\documentclass{ws-book975x65}

\usepackage{ws-book-har}           

\usepackage{tikz}

\begin{document}

\chapter[Control of chemical wave propagation]{Control of chemical wave propagation}

\author{Jakob L\"ober, Rhoslyn Coles, Julien Siebert, Harald Engel, Eckehard Sch\"oll}


\section{Introduction}

Besides the well-known Turing patterns, reaction-diffusion (RD) systems possess a rich variety of spatio-temporal structures
\cite{KUR84,mikhailov1990foundations,KAP95a}. 
Spatially one-dimensional examples include traveling fronts, solitary pulses, and periodic pulse trains that are the building blocks of more complicated patterns in two- and three-dimensional active media as, e.g., spiral and scroll waves, respectively.
Another important class of RD patterns forms stationary, breathing, moving or self-replicating localized spots.
Labyrinthine patterns as well as phase turbulence, defect-mediated spiral and scroll wave turbulence are examples for more complex patterns.
In the Belousov-Zhabotinsky (BZ) reaction in microemulsions the BZ inhibitor (bromide) is produced in nanodroplets and diffuses through the oil phase at a rate up to two orders of magnitude greater than that of the BZ activator (bromous acid).
In this heterogeneous RD system, a variety of patterns including three-dimensional Turing patterns have been observed by computer tomography (see \cite{bansagi2011tomography} and references therein).
\newline
Several control strategies have been developed for the purposeful manipulation of RD patterns.
Below, we will differentiate between closed-loop or feedback control with and without nonlocal spatial coupling \cite{DAH08,SCH09c,SIE14} or time delay \cite{KIM01,KYR09}, and open-loop control that includes external spatio-temporal forcing, optimal control \cite{troltzsch2010optimal}, control by imposed geometric constraints or heterogeneities, and others \cite{mikhailov2006control,schimansky2007analysis,vanag2008design,scholl2008handbook}.
While feedback control relies on continuously running monitoring of the system's state, open-loop control is based on a detailed knowledge of the system's dynamics and its parameters.
\newline
Feedback-mediated control has been applied quite successfully to the control of propagating one-dimensional (1D) waves as well as to spiral waves in 2D that are among the most prominent examples of spatio-temporal patterns in oscillatory and excitable RD systems.
Crucial for the control of spiral waves dynamics is the resonant drift of the spiral core in response to a periodic change of the medium's excitability exactly at the spiral's rotation frequency \cite{agladze1987observation}.
Under external resonant periodic forcing, the drift direction depends on the orientation of the spiral wave as there is no synchronization between the externally applied control signal and the intrinsic spiral wave dynamics.
In contrast, with feedback-mediated parametric forcing, the direction of resonant drift is independent of the spiral's current orientation.
Therefore, we can assign a unique vector to each point of the plane specifying the absolute value and the direction of feedback-induced resonant drift imposed on a spiral core located at the given position \cite{zykov2004feedback}. 
Stable fixed points and stable limit cycles are the possible attractors of the drift velocity field while separatrices of saddle points and unstable limit cycles form the boundaries of the basins of attraction.
Thus, from the drift velocity field we gain complete information about the asymptotic motion of the spiral core under feedback 
control.
Given that, first, feedback-induced drift remains small, and, second, the shape of the spiral wave can be approximated by an Archimedian spiral, from the RD equations for the concentration fields ordinary differential equations can be derived for the temporal variation of the core center coordinates. 
Different control loops have been realized in experiments with the photosensitive BZ reaction using feedback signals obtained from wave activity measured at one or several detector points, along detector lines, or in a spatially extended control domain including global feedback control.
Possible control parameters range from the gain and the time delay in the feedback loop over detector position to the size and geometrical shape of the control domain.
The theoretical predictions agree well with the experimental data, for details see for example \cite{schlesner2008efficient, zykov2004global}. 
Furthermore, feedback-mediated control loops can be employed in order to stabilize unstable patterns, such as unstable traveling wave segments and spots. This was shown in experiments with the photosensitive BZ reaction \cite{mihaliuk2002feedback}.
Two feedback loops were used to guide unstable wave segments along pre-given trajectories \cite{sakurai2002design}.
\newline
Under feedback control, a meandering spiral wave can be forced to rotate rigidly in a parameter range where rigid rotation is unstable in the absence of feedback 
\cite{schlesner2006stabilization}.
\newline
An open loop control was successfully deployed in dragging traveling chemical pulses of adsorbed CO during heterogeneous catalysis on platinum single crystal surfaces \cite{wolff2003gentle}.
In these experiments, pulse velocity was controlled by a laser beam creating a movable localized temperature heterogeneity on an addressable catalyst surface \cite{wolff2001spatiotemporal,wolff2002lokale,wolff2003wave}.
Dragging a chemical front or phase interface to a new position by anchoring it to a movable parameter heterogeneity, was studied theoretically in \cite{kevrekidis2004dragging,nistazakis2002targeted,malomed2002pulled}.
\newline
Many complex RD patterns can be understood as composed of interfaces, fronts, solitary excitation pulses etc. 
Before tackling control of complex patterns, it makes sense to develop a detailed understanding of the control of these simpler ``building blocks''.
We choose the Schl\"ogl model as a particularly simple, to some extent analytically tractable example of front dynamics in bistable RD media \cite{Schlogl1972crm, Schloegl1983fluctuations,SCH86a,SCH01}. 
In Sec. \ref{sec1}, experimentally feasible options are discussed for manipulating front dynamics. The effect of nonlocal feedback on front propagation is analyzed in Sec. \ref{sec2}, while Sec. \ref{sec3} presents an analytically derived open loop control tailored to a precise control of the front position over time.

\section{The Schl\"ogl model}\label{sec1}

\subsection{The Schl\"ogl model as an autocatalytic reaction mechanism}

\begin{figure}[ht]
  \centerline{
    \psfig{file=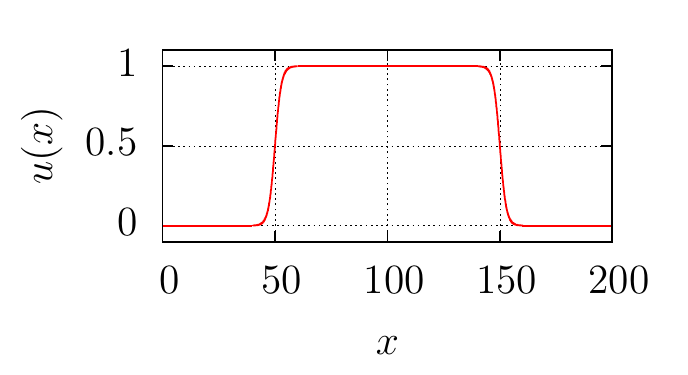,width=5cm}
    \psfig{file=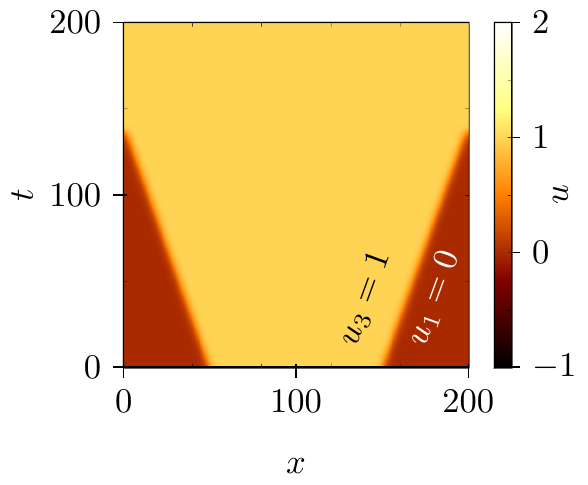,width=5cm}
  }
  \caption{Space-time plot of two traveling fronts (right) evolving from the symmetric initial concentration profile (left). 
  Numerical solution of the uncontrolled Schl\"ogl equations with periodic boundary conditions posed at the ends of a finite spatial domain. 
  Spatial discretization step: $\Delta x = 0.2$; maximum simulation time $T = 200$; time discretization step: $\Delta t = 0.01$. 
  Parameters $u_3 =1,\, u_2 = 0.25,\,u_1= 0,\,k=D=1$.}
  \label{fig.sieb.init}
\end{figure}

In 1972, Schl\"ogl discussed the autocatalytic trimolecular RD scheme \cite{Schlogl1972crm, Schloegl1983fluctuations}
\begin{align}
A_{1}+2X & \overset{k_{1}^{+}}{\underset{k_{1}^{-}}{\rightleftharpoons}}3X, & X & \overset{k_{2}^{+}}{\underset{k_{2}^{-}}{\rightleftharpoons}}A_{2}
\end{align}
as a prototype of a non-equilibrium first order phase transition. Concentrations $c_1$ and $c_2$ of the chemicals $A_1$ and $A_2$, respectively, are 
assumed to be kept fixed by appropriate feeding of the continuously stirred open reactor where the reaction takes place. Supply of $A_1$ and removal of $A_2$ maintains 
the RD system far from equilibrium. The reaction kinetics $R\left(u\right)$ of the chemical $X$ with concentration $u$ is cubic
\begin{align}
R\left(u\right) & =-k_{1}^{-}u{}^{3}+k_{1}^{+}c_{1}u{}^{2}-k_{2}^{+}u+k_{2}^{-}c_{2}.\label{eq:SchloeglReactionFunction}
\end{align}
Given that in some parameter range $R\left(u\right)=0$ possesses three real non-negative roots, $0\le u_1<u_2<u_3$, kinetics \eqref{eq:SchloeglReactionFunction} can be cast into the form
\begin{align}
R\left(u\right) &= -k\left(u-u_1\right)\left(u-u_2\right)\left(u-u_3\right)\label{eq:ReactionFunctionThreeRoots}
\end{align}
where the constant parameters $k$ and $u_i$ ($i = 1,2,3$) can be expressed by the parameters in $R\left(u\right)$, Eq. \eqref{eq:SchloeglReactionFunction}. Taking into 
account diffusion of $X$, the time evolution of the concentration field $u\left(x,t\right)$ is given by the RD equation
\begin{align}
\partial_t u & = D\partial^2_x u -k\left(u-u_1\right)\left(u-u_2\right)\left(u-u_3\right),\label{eq:ReactionDiffusionEquation}
\end{align}
where $D$ denotes the diffusion coefficient. Initially, Eq. \eqref{eq:ReactionDiffusionEquation} has been discussed in 1937 by Zeldovich and Frank-Kamenetsky in connection with 
flame propagation \cite{zeldovich1938theory}. Under spatially non-uniform conditions, concentrations $c_{1/2}$ are still assumed to be constant in space and time. 
For an open unstirred reactor fed solely by mass flow through the reactor boundaries, this 
assumption holds if components $A_{1/2}$ diffuse much faster than component $X$. A more sophisticated realization for e.g. microfluidic devices would be an application of spatially 
distributed nozzles within the reactor which continuously exchange the solution such that the concentrations of $A_{1/2}$ are kept constant everywhere.
A linear stability analysis reveals that $u_{1/3}$ and ${u_2}$ represent stable and unstable homogeneous steady states (HSS) of the RD system Eq. \eqref{eq:ReactionDiffusionEquation},
respectively. Therefore, in a certain parameter range the Schl\"ogl model describes a bistable chemical reaction. In the following, we focus on the control of the narrow interface between two coexisting domains of the stable and metastable phases $u_{1/3}$. Within this interface, concentration $u$ changes rapidly from one stable HSS to the other. 
The globally stable state tends to invade the entire available space developing an expanding or retracting front that displaces the metastable domain. For the Schl\"ogl model, 
the front profile $u\left(x,t\right) = U_c\left(z\right)$ in the comoving frame $z=x-ct$, and the front velocity, $c$, are known analytically
\begin{align}
U_c\left(z\right) &= \frac{1}{2}\left(u_1 + u_3\right) + \frac{1}{2}\left(u_1 - u_3\right)\tanh\left(\frac{1}{2}\sqrt{\frac{k}{2D}}\left(u_3-u_1\right)z\right),\label{eq:SchloeglTravelingFrontProfile}\\
c &= \sqrt{\frac{Dk}{2}}\left(u_1 + u_3 - 2u_2\right).\label{eq:SchloeglTravelingFrontVelocity}
\end{align}
The front solution $U_c\left(z\right)$ connects the upper stable stationary state $u_3$ for $z\rightarrow -\infty$ with the lower stable stationary state 
$u_1$ for $z\rightarrow\infty$. The front travels to the right as long as $u_1+u_3>2u_2$. A stationary front, i.e., a persistent spatial phase separation, 
exists only for the parameter combination with 
\begin{align}
u_1+u_3=2u_2 & \Rightarrow c=0 .\label{eq:StationaryFront}
\end{align}
Likewise, there exist front solutions which travel to the left with a negative velocity and satisfy interchanged boundary conditions.
Fig.\ref{fig.sieb.init} (right) shows a space-time plot of uncontrolled front propagation in response to the symmetric initial $u$-profile displayed in Fig.\ref{fig.sieb.init} (left).
Two fronts propagate at the same speed in opposite directions where the globally stable domain $u = u_3 $ (yellow) 
invades the metastable domain $u = u_1 $ shown in red in Fig.\ref{fig.sieb.init}.
\newline
In the following we write the controlled Schl\"ogl model in the general form 
\begin{align}
\partial_{t}u & =D\partial_{x}^{2}u+R\left(u\right) + F\left( u\right).
\label{eq:ControlledReactionDiffusionEquation}
\end{align}
Here, the control term $F$ is a functional of $u(x,t)$, which may contain time-delayed or nonlocal terms of the concentration variable 
$u$, described for instance by integrals over spatial or temporal kernels as discussed in the next section. 
The control term may also depend upon $x,t$, denoting an external spatio-temporal control signal. In case of open-loop control, the 
control signal is prescribed independently of the state of the system. For closed-loop or feedback control, the system state must be monitored, and $F$ turns out to be a functional of $u$ realizing a state-dependent control.
\newline
Experimentally, the state $u$ of the controlled system can be measured only with limited accuracy.
Often measurements are restricted to a certain region of the spatial domain supporting wave propagation. Furthermore, in multi-component RD systems rarely all components are simultaneously 
accessible to measurements. A system is called observable in control theory if it is possible to fully reconstruct its state from output measurements.
\newline
Subsequently we discuss different options for the control of the RD system Eq. \eqref{eq:ControlledReactionDiffusionEquation} in consideration of physically based constraints and derive the 
corresponding coupling functions.\\ 
First of all, feedback control (closed-loop control) may be described by nonlocal coupling with a kernel $G(x')$, e.g.,  
\begin{align}
F(u)=\int_{-\infty}^{+\infty} G(x')u(x-x')dx'
\end{align}
or distributed time-delayed feedback with a kernel $G(t')$
\begin{align}
F(u)=\int_{0}^{+\infty} G(t')u(t-t')dt'.
\end{align}

Second, an additive external forcing $f\left(x,t\right)$ that describes time-dependent local concentration sources or sinks leads to $F=f(x,t)$ (open-loop control). In this case $f$ specifies the current 
local rate at which component $X$ is added to or removed from a reactor.
An experimental realization is a nearly continuous array of nozzles diluting or concentrating the solution and thereby changing the concentration of component $X$ at the prescribed rate $f$. The rate 
can be positive or negative and is not subject to any explicit restrictions. 
However, $X$ cannot be removed if its concentration $u$ reaches its minimum possible value $u=0$. This state constraint applies to any RD system whose components represent chemical concentrations.
\newline
Further, control can be realized via parameters in the reaction kinetics. We set $F=\mathcal{G}\left( u\right)f\left(x,t\right)$
where $\mathcal{G}$ is a possibly state-dependent function that depends on the mechanism by means of which the control signal is coupled to the RD system. If the reaction function $R\left(u;p\right)$ depends on some experimentally controllable parameter 
$p$, we replace this parameter by $p +\epsilon f\left(x,t\right)$. Expanding with respect to the small control amplitude $\epsilon$ it follows in leading order
\begin{align}
\mathcal{G}\left( u\right) & = \partial_p R\left(u;p\right).\label{eq:CouplingFunction}
\end{align}
Assume, for example, that concentration $c_2$ in the Schl\"ogl model can be controlled spatio-temporally. Then, replacing the constant $c_2$ in Eq. \eqref{eq:SchloeglReactionFunction} by the space-time dependent 
quantity $c_2 + \epsilon f\left(x,t\right)$, along the same line of reasoning as before we end up with a state-independent coupling 
\begin{align}
\mathcal{G} & = k_{2}^{-}.
\end{align}
However, now the explicit control constraint $c_2+\epsilon f\left(x,t\right)> 0$ has to be valid. A similar consideration starting with concentration $c_1$ leads to a state constraint 
$c_1 +\epsilon f\left(x,t\right)>0$ and a multiplicative control with state-dependent coupling function 
\begin{align}
\mathcal{G} & =k_{1}^{+}u{}^{2}.
\end{align}
Finally, apart from the concentrations $c_{1/2}$, the reaction rate $R\left(u\right)$ Eq. \eqref{eq:SchloeglReactionFunction} depends on the reaction coefficients $k_{1/2}^{\pm}$ that in turn are 
functions of the temperature $T$ according to the Arrhenius law $k_{1/2}^{\pm}\sim e^{-E_{1/2}^{\pm}/\left(k_{B}T\right)}$. Here, $k_B$ and $E_{1/2}^{\pm}$ denote the Boltzmann constant and activation 
energy of the corresponding reaction. The temperature dependence was exploited for the control of pattern formation in the catalytic CO oxidation on a single crystal Pt(110) surface 
\cite{wolff2001spatiotemporal}. In the experiments, a computer-controlled movable laser beam induced a localized temperature heterogeneity from which reaction fronts and pulses nucleated 
\cite{wolff2003wave}. Gently dragged by the laser beam, pulses were moved over the surface with a velocity up to twice as large as the velocity of the uncontrolled pulse \cite{wolff2003gentle}. 
The temperature field response is very fast compared to the time scale of the catalytic reaction, and the heat released by the reactions is negligible. 
Modeling the controlled reaction by three coupled RD equations with a laser-induced localized Gaussian temperature heterogeneity of small amplitude ($\left|T\left(x,t\right)\right| \leq 1K$) 
and small standard deviation ($2\mu \text{m}$) qualitatively confirmed the experimental results.
To incorporate a temperature-mediated control of the Schl\"ogl model Eq. \eqref{eq:ControlledReactionDiffusionEquation},
we substitute $T$ by $T+\epsilon f\left(x,t\right)$ and obtain for small $\epsilon$ the coupling function
\begin{align}
\mathcal{G}\left(u\right) & =\frac{1}{k_{B}T^{2}}\left(E_{1}^{-}k_{1}^{-}u^{3}+E_{2}^{+}k_{2}^{+}u-c_{1}E_{1}^{+}k_{1}^{+}u^{2}-c_{2}E_{2}^{+}k_{2}^{-}\right).
\end{align}
Assuming for simplicity that all activation energies are equal, $E_{1}^{-}=E_{2}^{-}=E_{1}^{+}=E_{2}^{+}=E$, the coupling function turns out to be proportional to the reaction rate $R$,
\begin{align}
\mathcal{G}\left(u\right) & =-\frac{E}{k_{B}T^{2}}R\left( u\right)
\end{align}
and the controlled Schl\"ogl model reads
\begin{align}
\partial_{t}u & =D\partial_{x}^{2}u+R\left(u\right) -\frac{\epsilon E}{k_{B}T^{2}}R\left(u\right)f\left(x,t\right)
\end{align}
subjected to the constraint $T+\epsilon f\left(x,t\right)>0$.
\newline
For the Schl\"ogl model with a coupling function $\mathcal{G} \left(u\right) \sim R\left(u\right)$, the effect of a stationary periodic modulation $f\left(x\right)$ on the propagation velocity 
of traveling front solutions was analytically investigated in \cite{alonso2010wave,loeber2012front}. In accordance with numerical simulations, a spatially 
periodic modulation with zero spatial average generally lowers the average propagation velocity. Propagation failure, or pinning, of fronts occurs if the spatial period of the modulation 
is of the same order as the front width. Outside this interval, i.e., for smaller and larger periods, front propagation is still possible. Thus, in a proper range of 
spatial periods a temperature modulation results in a persistent spatial phase separation.
\newline

Finally, we briefly mention control by non-uniform boundary conditions. This method, not encompassed by the chosen form of a controlled RD system according to Eq. \eqref{eq:ControlledReactionDiffusionEquation},
is nevertheless important for applications. Boundary control assumes that the mass exchange flow transfer $J\left(t\right)$ of chemical species $X$ can be prescribed at the boundaries such that the boundary 
conditions in the one-dimensional finite domain $\left[x_0,x_1\right]$ are given by 
\begin{align}
\partial_x u\left(x_0, t\right) & = J_0\left(t\right),&\partial_x u\left(x_1, t\right) &= J_1\left(t\right).
\end{align}
with continuously adjustable boundary flows $J_{0/1}\left(t\right)$. We refer to \cite{PhysRevLett.91.208301} for an example of how a desired stationary concentration profile can be enforced onto a 
RD system applying boundary control. More difficult approaches based on control signals coupled nonlinearly into the system are possible and experimentally relevant but not considered in this 
contribution.

\subsection{Experimental realizations of the Schl\"ogl model}

Examples of isothermal chemical systems exhibiting multiple stationary states are few. The Schl\"ogl model does not describe
a realistic reaction scheme because it involves a trimolecular reaction step, and the 
latter is based on the unlikely reactive collision between three molecules within a small volume. However, due to Korzhukin's theorem, for any homogeneous chemical reaction with polynomial 
reaction rate one can write down an equivalent set of unimolecular reactions between intermediate species \cite{erdi1989mathematical, Korzukhin1967mathematicala, Korzukhin1967mathematicalb}.
One notable example is the iodate oxidation of arsenous acid \cite{papsin1981bistability}. The resulting reaction rate in a RD equation Eq. \eqref{eq:ReactionDiffusionEquation} for the 
concentration of iodide ions $u = \left[\text{I}^-\right]$ reads \cite{hanna1982detailed}
\begin{align}
R\left( u\right) & = \left(k_1 + k_2 u\right)\left(c_1 -u\right)c_2u.
\end{align}
Here, $c_1=\left[\text{IO}^-_3\right]$ denotes the concentration of iodate while $c_2=\left[ \text{H}^+\right]^2$ where $\left[ \text{H}^+\right]$ is the concentration of hydrogen ions.
In the uncontrolled system concentrations $c_1$ and $c_2$ are assumed to be constant. Obviously, both of them, as well as the rate constants $k_{1/2}$ and the temperature could be considered 
for spatio-temporal control as described above in the case of the Schl\"ogl model.
\newline
Another elaborately studied experimental example for chemical bistability is the CO oxidation on Pt(111) single crystal surfaces. This reaction has been modeled by a two coupled RD equations 
for the surface concentrations of adsorbed CO and oxygen. In this case, bistability relies on the Langmuir-Hinshelwood mechanism. Possible control parameters are partial pressures of CO and oxygen 
in the gas phase \cite{bar1992reaction}. 
Front propagation has been also observed in experiments with the CO oxidation on Pt(110) \cite{nettesheim1993reaction}. 
\newline
Because of the close similarity of generation and recombination processes of charge carriers in semiconductors with chemical reactions,
the Schl{\"o}gl model can also be applied to self-organization in semiconductors induced by nonlinear generation and recombination
processes \cite{SCH87,SCH01}. The analogy of pattern formation with chemical and electrochemical systems 
\cite{PLE01}, and in particular the front dynamics in bistable semiconductor models \cite{MEI00b}, and its control by
time-delayed feedback \cite{KEH09,SCH09} has been extensively studied. 
\newline
Another non-chemical model of bistable dynamics, which is nevertheless quite interesting from the viewpoint of spatio-temporal control, is a liquid crystal light valve (LCLV) inserted in an optical feedback 
loop \cite{Residori2005201, haudin2010front, haudin2009driven}. A nematic liquid crystal film is placed between a glass plate and a photoconductive material. Applying an external voltage $V_0$ across 
the cell with the help of transparent electrodes orients the polar molecules in parallel to the electric field. A spatio-temporal illumination distribution on the photoconductive wall modulates 
the electric field locally and allows for spatio-temporal control of the orientation of the liquid crystal molecules. Near to the so-called point of "nascent bistability`` the normalized average 
director $u\left(x,t\right)$ obeys the RD equation
\begin{align}
\partial_t u &= D\partial_x^2 u - u^3 + \tilde{\epsilon} u + \eta +\left(b+du \right)f\left(x,t\right).\label{eq:LCLVBistableSystem}
\end{align}
Here, $f$ denotes the spatio-temporal forcing signal that is proportional to the applied light intensity, while constants $\eta,\,\tilde{\epsilon},\,b$ and $d$ are related to the properties of the LCLV and 
depend on the applied voltage. Investigation of fronts propagating through a periodically modulated medium revealed the existence of a pinning range. Instead of a single parameter combination leading 
to a stationary front, Eq. \eqref{eq:StationaryFront}, a whole range of parameters results in a stationary phase separation \cite{haudin2009driven}.\\
\newline
Finally, we note that all controlled one-component models with cubic nonlinearity can be rescaled and cast in a particularly simple form. This can be seen as follows. Starting from the
cubic reaction rate $R\left(u\right)$ expressed in terms of its three roots, Eq. \eqref{eq:ReactionFunctionThreeRoots}, we introduce
a rescaled concentration $U$, a new parameter $\alpha$ and rescaled space and time scales according to
\begin{align}
U &= \frac{u-u_1}{u_3-u_1}, & t &= \frac{T}{k\left(u_3-u_1\right)^2}, & 
x &= \frac{X}{\left(u_3-u_1\right)} \sqrt{\frac{D}{k}}, & \alpha = \frac{u_2-u_1}{u_3-u_1}.
\end{align}
Now, the rescaled Schl\"ogl model reads
\begin{align}
\partial_T U &= \partial_X^2 U + \tilde{R}\left(U\right) + \tilde{\mathcal{G}}\left(U\right)\tilde{f}\left(X,T\right).\label{eq:RescaledControlledSchloeglModel}
\end{align}
The corresponding reaction function $\tilde{R}$, traveling front solution $\tilde{U}_c$, and velocity $\tilde{c}$
are obtained by substituting $u_1\rightarrow 0,\,u_2\rightarrow \alpha,\,u_3 \rightarrow 1,\,D\rightarrow 1$ and $k \rightarrow 1$ in 
Eq. \eqref{eq:SchloeglTravelingFrontProfile}, Eq. \eqref{eq:SchloeglTravelingFrontVelocity} and Eq. \eqref{eq:ReactionFunctionThreeRoots}, respectively.
This gives finally
\begin{align}
\tilde{R}\left(U\right) &= -U\left(U-\alpha\right)\left(U-1\right),\label{eq:RescaledReactionFunction}\\
\tilde{U}_c\left(x\right) &= \frac{1}{1+e^{\frac{x}{\sqrt{2}}}},\label{eq:RescaledSchloeglTravelingFrontProfile}\\
\tilde{c} &= \frac{1}{\sqrt{2}}\left(1-2\alpha\right).\label{eq:RescaledFrontVelocity}
\end{align}
Note that the coupling function is modified under rescaling, too,
\begin{align}
\frac{1}{k\left(u_3-u_1\right)^3}\mathcal{G}\left(u\right) &=\frac{1}{k\left(u_3-u_1\right)^3}\mathcal{G}\left(\left(u_3-u_1\right)U + u_1\right) =\tilde{\mathcal{G}}\left(U\right),
\end{align}
however, qualitative changes do not occur: a constant $\mathcal{G}$ stays constant under rescaling, a polynomial of degree $n$ in $\mathcal{G}$ stays the same in $\tilde{\mathcal{G}}$, etc. The control signal 
$f\left(x,t\right) = ~ f\left(X,T\right)$ is now expressed in the rescaled coordinates.
We will use the rescaled Schl\"ogl equation \eqref{eq:RescaledControlledSchloeglModel} for all numerical simulations in the following chapters.

\newcommand{\sinc}[1]{\operatorname{sinc}\left( #1 \right)}

\section{Nonlocal control of space-time patterns in the Schl\"ogl model}\label{sec2}

\subsection{Feedback control of the front propagation}

We consider feedback control of the reaction-diffusion system by adding either a distributed nonlocal feedback term 
(Eq.\ref{Eq.nl}) or a distributed time-delayed feedback (Eq.(\ref{Eq.td})):
\begin{eqnarray}
  \partial_t u &=&  R(u)  + \partial^2_x u + \sigma\left[ \int_{-\infty}^{+\infty} G(x')u(x-x')dx' - u(x) \right]
  \label{Eq.nl} \\
  \partial_t u &=&  R(u)  + \partial^2_x u + \sigma\left[ \int_{0}^{+\infty} G(t')u(t-t')dt' - u(t) \right]
  \label{Eq.td}
\end{eqnarray}
where $G(x)$ and $G(t)$ are normalized integral kernels whose integrals equal unity. Here we have chosen a form of the coupling
term which is non-invasive for homogeneous or stationary states, respectively, i.e., it vanishes if $u(x,t)$ does not depend upon space (Eq.(\ref{Eq.nl})) or time (Eq.(\ref{Eq.td})). If $G(t)=\delta(t-\tau)$, Eq.(\ref{Eq.td}) reduces to Pyragas time-delayed feedback
control \cite{PYR92}. Table \ref{tab.ker} gives an overview of the investigated kernels.

The motivation for these distributed feedback forms comes from the elimination of one equation in the following two-variable reaction-diffusion system:
\begin{eqnarray}
\partial_t u &=&  R(u)  - g w  + \partial^2_x u,  \label{Eq.u} \\  
\tau \partial_t w &=&  h u - f w + D_w \partial^2_x w , \label{Eq.w}
\end{eqnarray}
where all the terms are linear except the function $R(u)$, see Eq.\eqref{eq:RescaledReactionFunction}. The two concentrations $u$ and $w$ correspond, respectively, to the activator and the inhibitor, and are linearly coupled by the terms $-gw$ and $hu$. The parameter $D_w$ is the inhibitor diffusion coefficient ($D_w>0$). 
The distributed nonlocal feedback term is obtained by the elimination of the second equation Eq.(\ref{Eq.w}) in the limit when $\tau \rightarrow 0$ \cite{KAN02,NIC02,SHI04,COL13,COL14,SIE14}. The time-delayed feedback is obtained by eliminating the second equation in the limit when $D_w\rightarrow 0$.

\begin{table}[!htbp]
  \tbl{Kernels and their respective Fourier and Laplace transforms. (a) Nonlocal symmetric kernels. (b) Nonlocal asymmetric kernels. (c) Time-delayed kernels.}
  {
    \begin{tabular}{@{}cccc@{}}
      \toprule
      (a) & Name & $G(x)$ & Fourier transform $\mathcal{F}\{G\}(k)$\\
      \colrule
      & centered exponential & $\dfrac{1}{2a}e^{-\tfrac{|x|}{a}}$ & $\dfrac{1}{\sqrt{2\pi}}\cdot\dfrac{1}{1 + (ak)^2}$ \\[12pt]
      
      & symmetric exponential$^{\text a}$ & $\dfrac{1}{4a}\left(e^{-\tfrac{|x-d|}{a}} + e^{-\tfrac{|x+d|}{a}}\right)$ & $\dfrac{\cos{dk}}{\sqrt{2\pi}}\cdot\dfrac{1}{1 + (ak)^2}$ \\[12pt]
      
      & Gaussian & $\dfrac{1}{a\sqrt{2\pi}}e^{-\tfrac{x^2}{2a^2}}$ & $\dfrac{1}{\sqrt{2\pi}}e^{\tfrac{-k^2a^2}{2}} $ \\[12pt]
      
      & Mexican hat & $\dfrac{1}{a}\sqrt{\dfrac{2}{\pi}}\left(1-\dfrac{x^2}{2a^2}\right) e^{-\tfrac{x^2}{2a^2}}$ & $\dfrac{2}{\sqrt{2\pi}}(1 + \dfrac{a^2k^2}{2}))e^{\dfrac{-a^2 k^2}{2}}$\\[12pt]
      
      & centered rectangular & $a\Pi(ax)$ & $\dfrac{1}{\sqrt{2 \pi}}\cdot \sinc{\dfrac{k}{2\pi a}}$ \\[12pt]
      
      & symmetric rectangular$^{\text b}$ & $\dfrac{a}{2}\left(\Pi(a(x+d)) + \Pi(a(x-d))\right)$ & $\dfrac{\cos{dk}}{\sqrt{2 \pi}}\cdot \sinc{\dfrac{k}{2\pi a}}$ \\[12pt]
      \botrule
      \toprule
      (b) & Name & $G(x)$ & Fourier transform $\mathcal{F}\{G\}(k)$\\
      \colrule
      & shifted exponential & $\dfrac{1}{2a}e^{-\tfrac{|x-d|}{a}}$ & $\dfrac{e^{-idk}}{\sqrt{2\pi}}\cdot\dfrac{1}{1 + (ak)^2}$ \\[12pt]
      
      & shifted Gaussian & $\dfrac{1}{a\sqrt{2\pi}}e^{-\tfrac{(x-d)^2}{2a^2}}$ & $\dfrac{e^{-idk}}{\sqrt{2\pi}}e^{\tfrac{-k^2a^2}{2}} $ \\[12pt]
      
      & shifted rectangular & $a\Pi(a(x-d))$ & $\dfrac{e^{-idk}}{\sqrt{2 \pi}}\cdot \sinc{\dfrac{k}{2\pi a}}$ \\[12pt]
      \botrule
      \toprule
      (c) & Name & $G(t)$ & Laplace transform $\mathcal{L}\{G\}(\lambda)$\\
      \colrule
      & uniform delay$^{\text c}$ & $a\Pi(a(t-\tau))$ & $\dfrac{2a}{\lambda}\sinh{a\lambda/2}e^{-\lambda\tau} $ \\[12pt]
      
      & weak gamma delay & $ae^{-at}$ & $\dfrac{a}{a+\lambda} $ \\[12pt] 
      
      & strong gamma delay & $a^2t e^{-at}$ & $\dfrac{a^2}{(a+\lambda)^2} $ \\[12pt]
      \botrule
    \end{tabular}
  }
  \begin{tabnote}
    $^{\text a}$ With $d/a >> 1$, since $\int G(x)\mathrm{d}x = 1 - \tfrac{1}{2} e^{-d/a}$.
  \end{tabnote}
  \begin{tabnote}
    $^{\text b}$ With $d - \dfrac{1}{2a} > 0$.
  \end{tabnote}
  \begin{tabnote}
    $^{\text c}$ With $\dfrac{a}{2} < \tau$.
  \end{tabnote}

  \label{tab.ker}
\end{table}

\subsection{Stability analysis of the homogeneous steady states}
From Eq.(\ref{Eq.nl}) and Eq.(\ref{Eq.td}), dispersion relations Eq.(\ref{Eq.nl.lambda}) and Eq.(\ref{Eq.td.lambda}), respectively, are obtained by performing a linear stability analysis around the homogeneous steady state solutions $u_1,\,u_2,\,u_3$ by setting $u(x,t) = u_{1,2,3} + \delta u$ with small $\delta u = e^{-ikx}e^{\lambda t}$:

\begin{figure}[ht]
  \centerline{
    \psfig{file=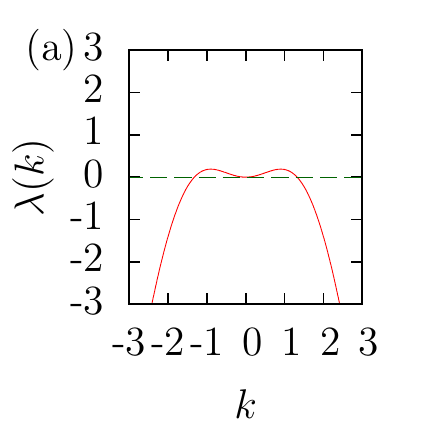,width=4cm}
    \psfig{file=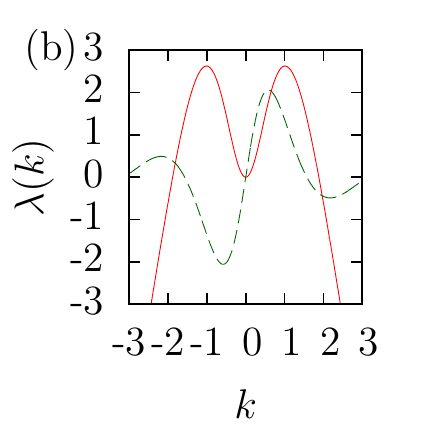,width=4cm}
    \psfig{file=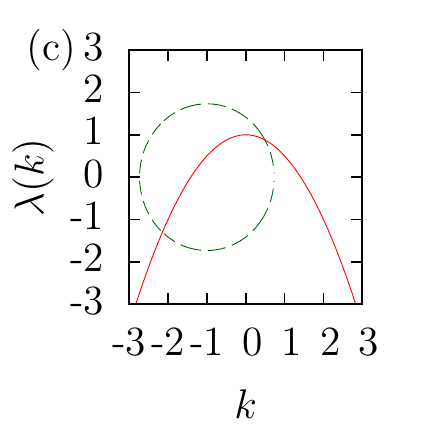,width=4cm}
  }

  \caption{Dispersion relation of the homogeneous steady state $u_1$. (solid) Re$\lambda(k)$. (dashed) Im$\lambda(k)$. Obtained via (a) Eq.(\ref{Eq.nl.lambda}) with (a) the nonlocal symmetric kernel $G(x)=1/(2a) exp(|x|/a)$, (b) the nonlocal asymmetric kernel $G(x)=1/(2a) exp(|x-d|/a)$, and (c) Eq.(\ref{Eq.td.lambda}) with the time-delayed kernel $G(t) = a exp(-at)$. Parameters $\alpha=0$, $a=1$, $d=2$ and $\sigma = -3$.}
  \label{fig.sieb.dr}
  
\end{figure}

\begin{eqnarray}
  \lambda &=& R'(u^*) - k^2 + \sigma(\sqrt{2\pi}\mathcal{F}\{G\}(k) - 1), \label{Eq.nl.lambda} \\
  \lambda &=& R'(u^*) - k^2 + \sigma(\mathcal{L}\{G\}(\lambda) - 1), \label{Eq.td.lambda}
\end{eqnarray}
where $\mathcal{F}\{G\}(k)$ and $\mathcal{L}\{G\}(\lambda)$ are the Fourier transform of the kernel $G(x)$ and the Laplace transform of the kernel $G(t)$, respectively.

\begin{figure}[htb]
  \centerline{
    \psfig{file=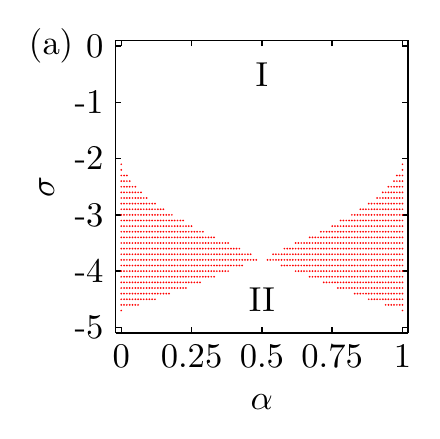,width=4cm} 
    \psfig{file=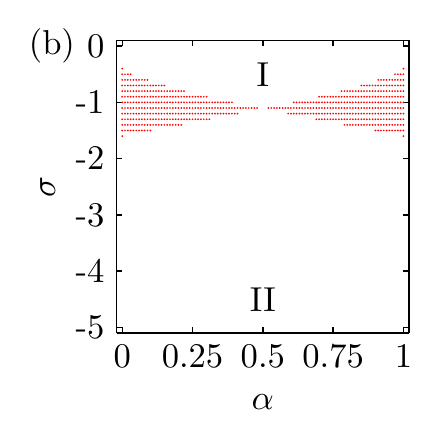,width=4cm} 
    \psfig{file=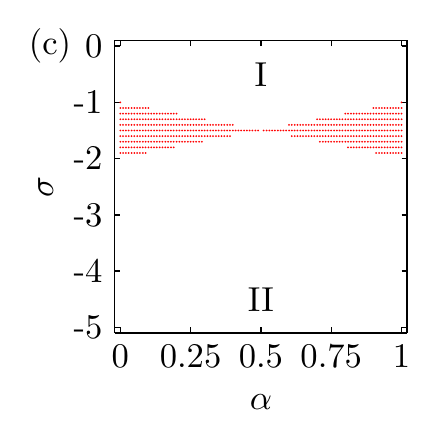,width=4cm} 
  }
  \caption{Stability of the homogeneous steady states ${u_1,u_3}$ in the $(\sigma,\alpha)$ parameter plan. 
  Region I: both homogeneous steady states are stable \textit{i.e.} Re$\lambda(k)<0$ for both ${u_1,u_3}$. (dotted area): 
  one of the homogeneous steady state is unstable while the other is stable. Region II: both homogeneous steady states are 
  unstable \textit{i.e.} Re$\lambda(k)\geq 0$ for both ${u_1,u_3}$. Instability are: (a) Turing instability, (b) traveling 
  waves instability and (c) Hopf instability. Results obtained via Eq.(\ref{Eq.nl.lambda}) with (a) the nonlocal symmetric kernel 
  $G(x)=1/(2a) exp(|x|/a)$, (b) the nonlocal asymmetric kernel $G(x)=1/(2a) exp(|x-d|/a)$ and (c) Eq.(\ref{Eq.td.lambda}) the 
  time-delayed kernel $G(t) = a exp(-at)$. Parameters $a=1$, $d=2$. }
  \label{fig.sieb.dr.as}
\end{figure}

From these dispersion relations, one can see that different kinds of instabilities may appear for suitable choice of $\sigma$: 
(i) nonlocal symmetric kernels may lead to a Turing instability (Im$(\lambda)=0$, Re$(\lambda)>0$ for finite wave number $k\neq0$), 
(ii) nonlocal asymmetric kernels may lead to a traveling wave instability (Im$(\lambda)\neq0$, Re$(\lambda)>0$ for finite wave number $k\neq0$) and 
(iii) time-delayed kernel may lead to a Hopf instability (Im$(\lambda)\neq0$, Re$(\lambda)>0$ for wave number $k=0$).
Fig.\ref{fig.sieb.dr} illustrates dispersion relations for these three different kinds of kernels.
Furthermore, solving Re~$\lambda(k) \geq 0$ leads to the phase diagrams of instabilities of the homogeneous steady states shown in Fig.\ref{fig.sieb.dr.as}.

\subsection{Simulations}
Numerical simulations of Eqs.(\ref{Eq.nl}) and (\ref{Eq.td}) have been performed. The behavior of the system can be classified into four types depending on the value of the coupling strength $\sigma$.

\begin{figure}[!htbp]
  \sidebyside{
     (a)\hspace*{5cm}
  }{
     (b)\hspace*{5cm}
  }
  \sidebyside{
    \psfig{file=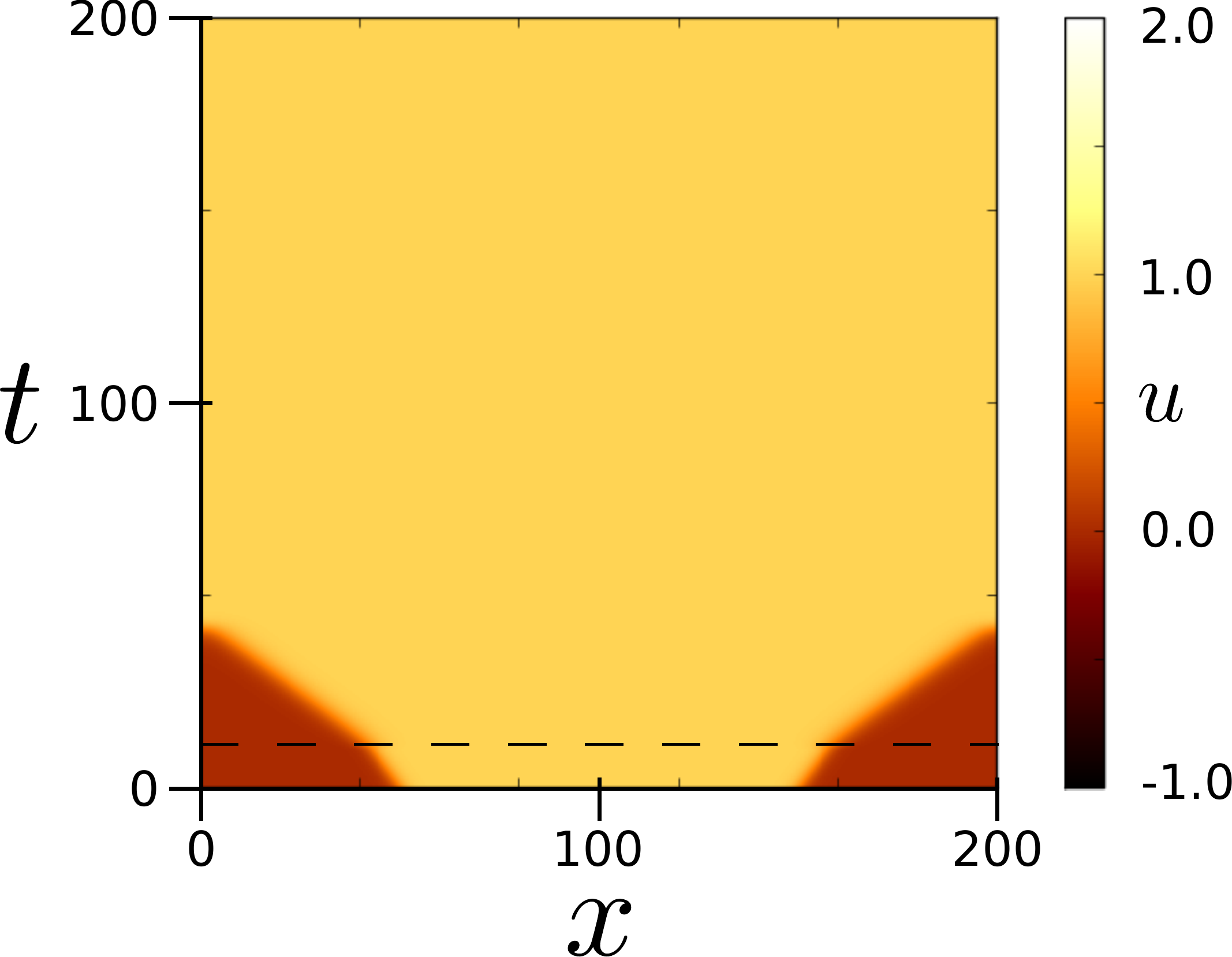,width=5cm}
  }{
    \psfig{file=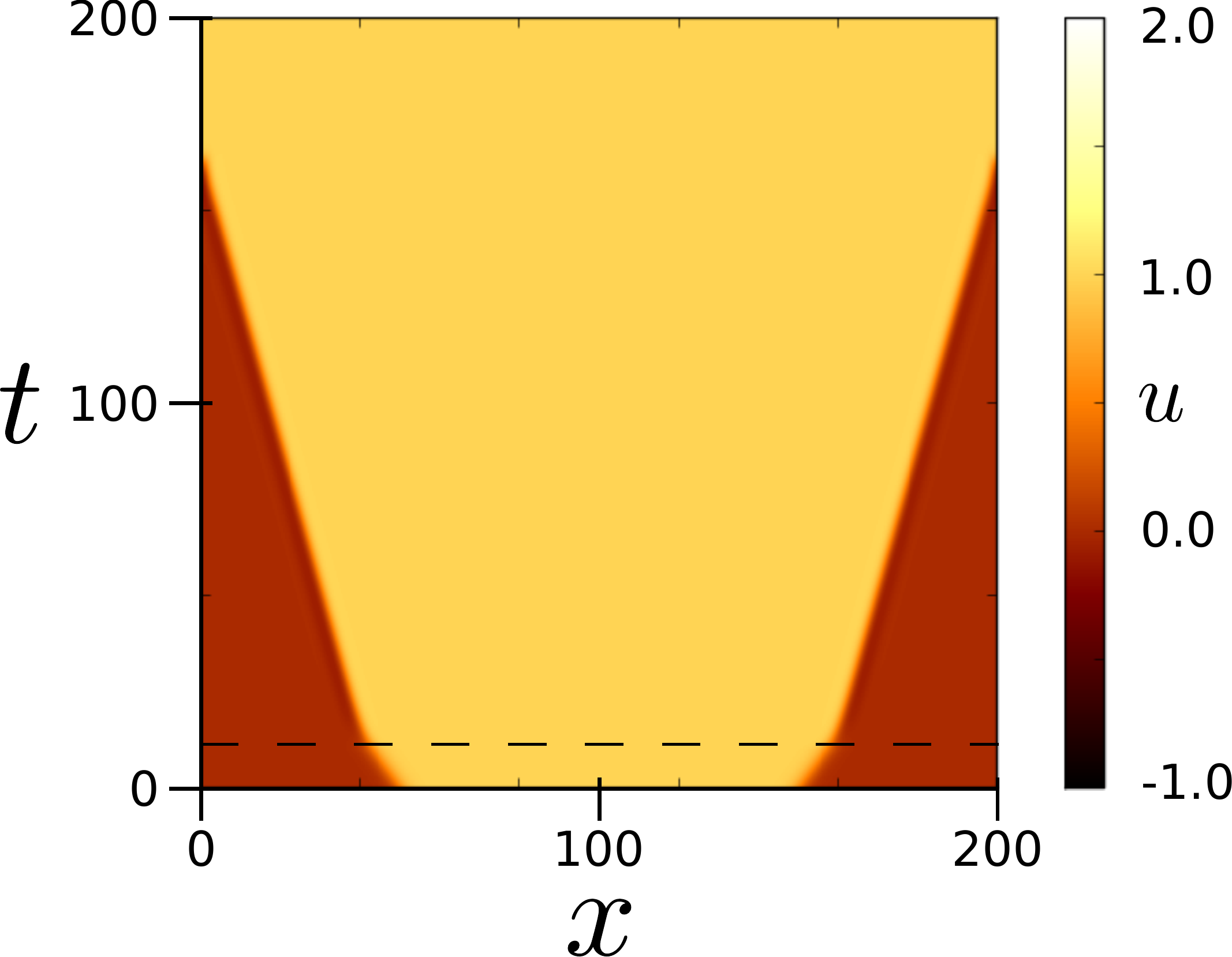,width=5cm}
  }

  \sidebyside{
     (c)\hspace*{5cm}
  }{
     (d)\hspace*{5cm}
  }
  \sidebyside{
    \psfig{file=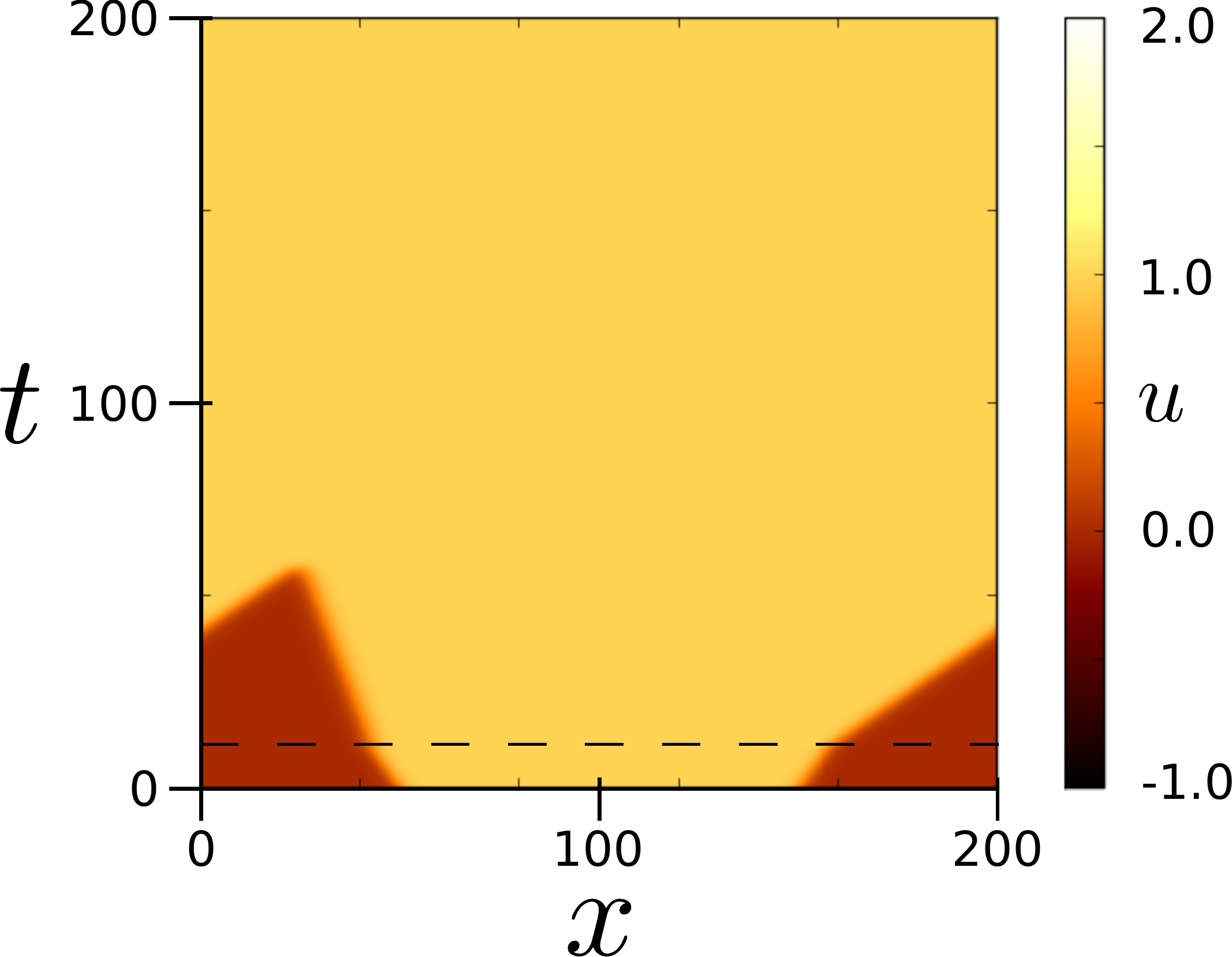,width=5cm}
  }{
    \psfig{file=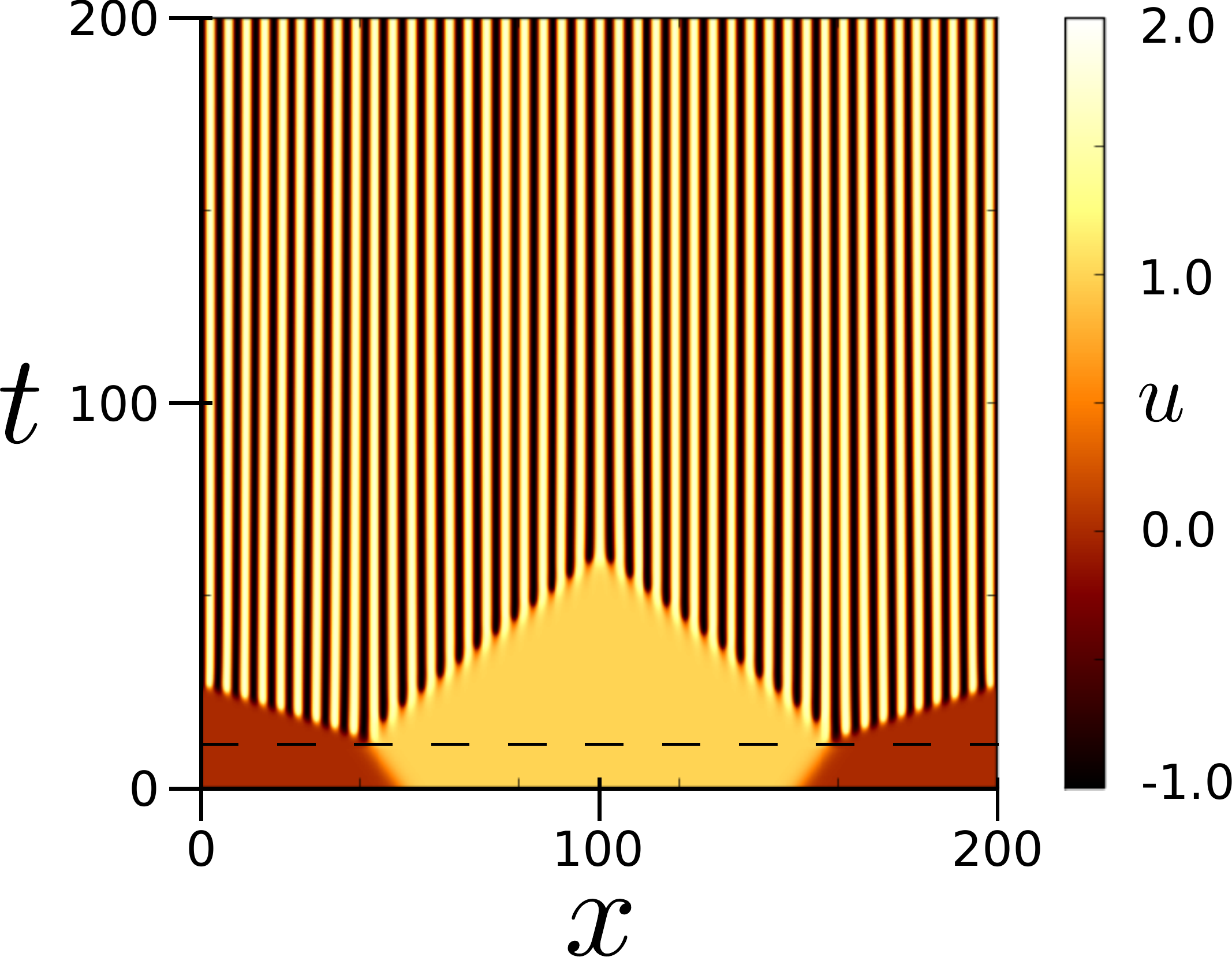,width=5cm}
  }

  \sidebyside{
     (e)\hspace*{5cm}
  }{
     (f)\hspace*{5cm}
  }
  \sidebyside{
    \psfig{file=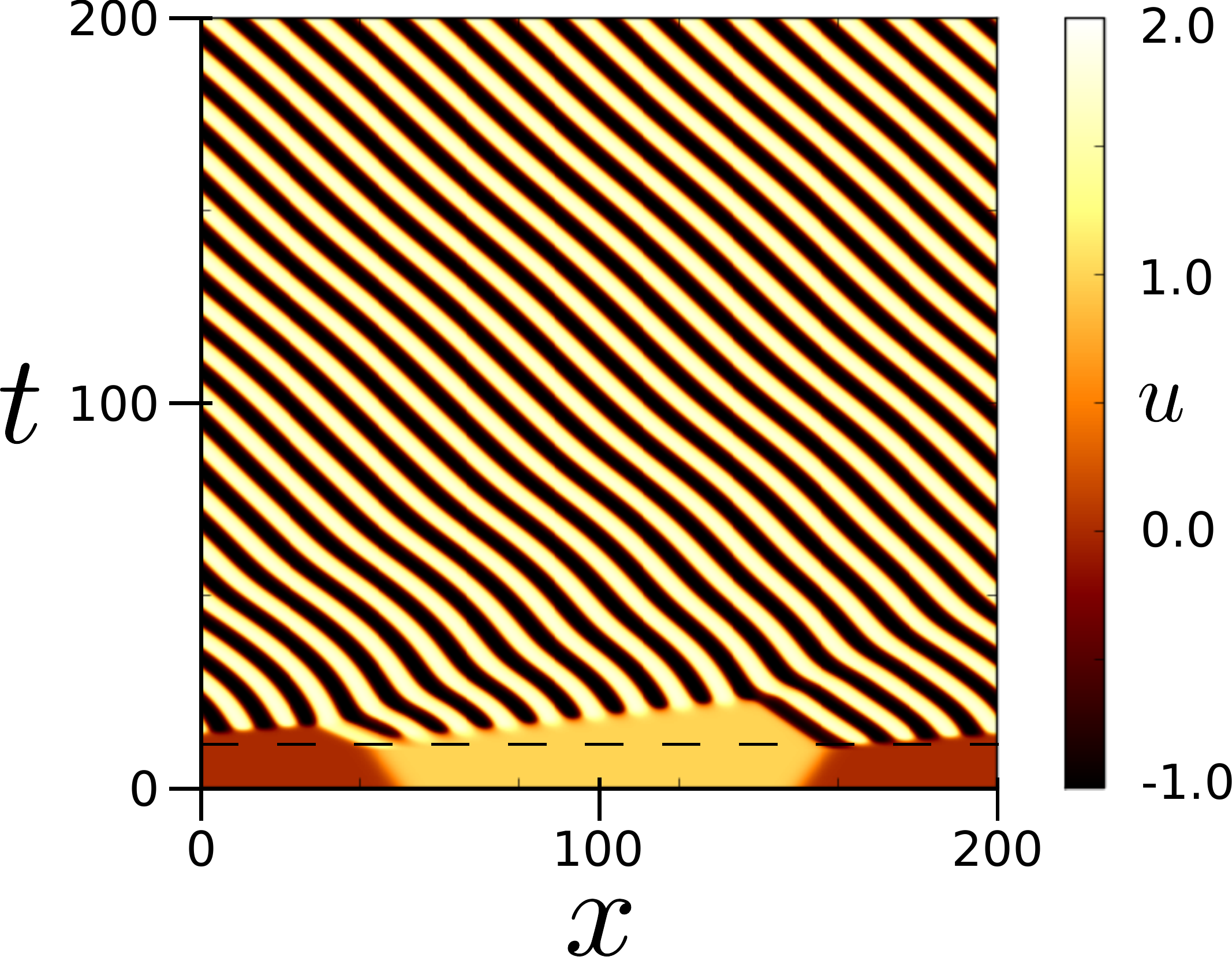,width=5cm}
  }{
    \psfig{file=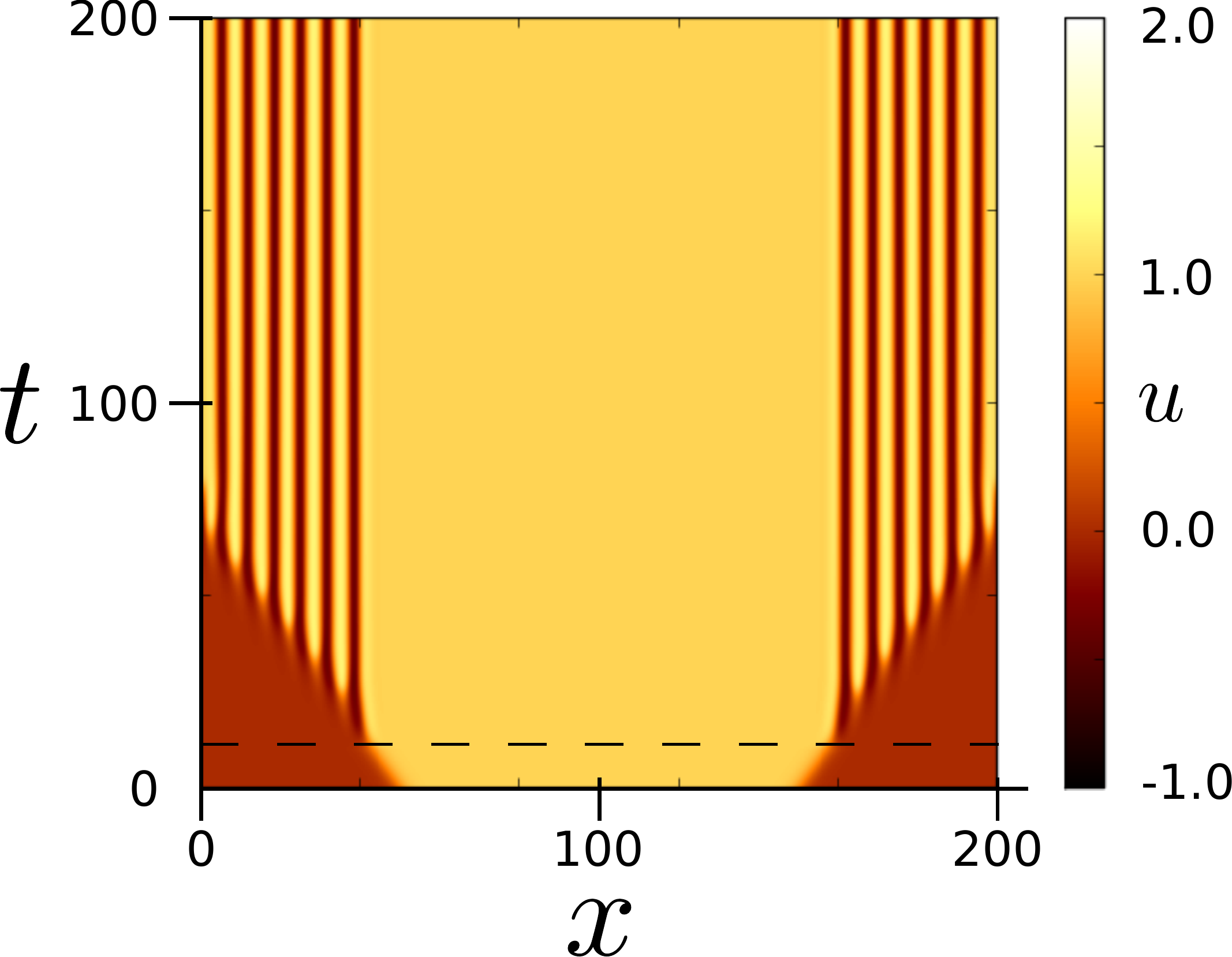,width=5cm}
  }
  
  \sidebyside{
     (g)\hspace*{5cm}
  }{
  }
  \sidebyside{
    \psfig{file=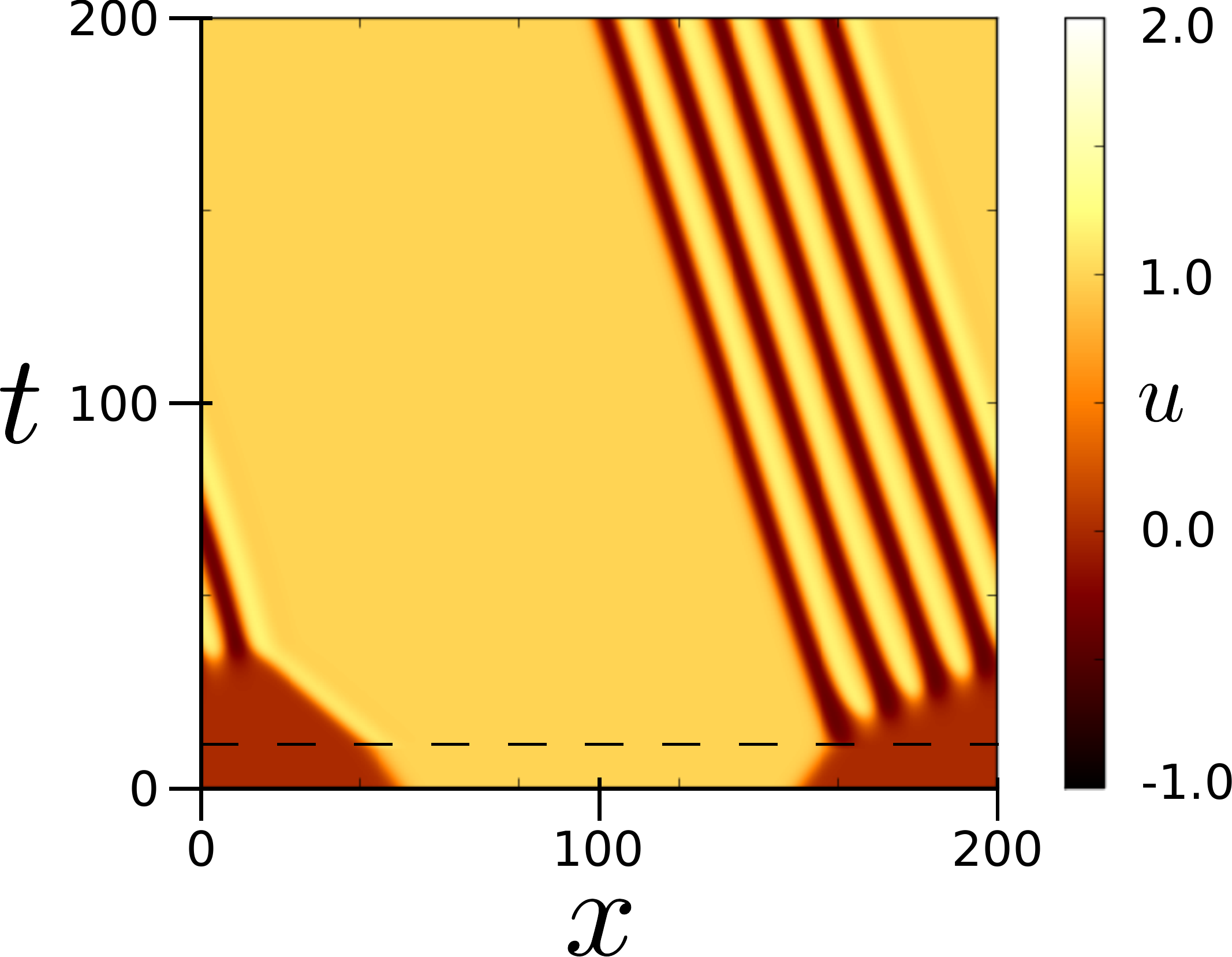,width=5cm}
  }{
  }
  
  \caption{Space-time patterns for distributed nonlocal feedback: (a) acceleration of the fronts; (b) deceleration of the fronts; (c) asymmetric acceleration; (d) Turing patterns; (e) traveling waves; (f) coexistence of Turing patterns and homogeneous state; (g) coexistence of traveling waves and homogeneous state. Results obtained via numerical simulation of Eq.\ref{Eq.nl} with $\alpha=0$. (a--d) Gaussian kernel $G(x)= 1/(a\sqrt{2\pi}) \exp(-x^2/(2a^2))$, $a=1$; (a) $\sigma=5$; (b) $\sigma=-2$; (c) $\sigma=-3$; (d) $\sigma=-5$. (e--g) Shifted exponential kernel $G(x)=1/(2a) \exp(-|x-d|/a)$, $a=1$, $d=5$; (f) $\sigma=0.25$; (g) $\sigma=-1.0$; (h) $\sigma=-0.25$.
  Time scale (a--g) $0\leq t\leq 200$, space scale (a--g) $0\leq x\leq 200$. }
  \label{fig.sieb.nl}
\end{figure}

\begin{figure}[!htbp]
  \sidebyside{
     (a)\hspace*{5cm}
  }{
     (b)\hspace*{5cm}
  }
  \sidebyside{
    \psfig{file=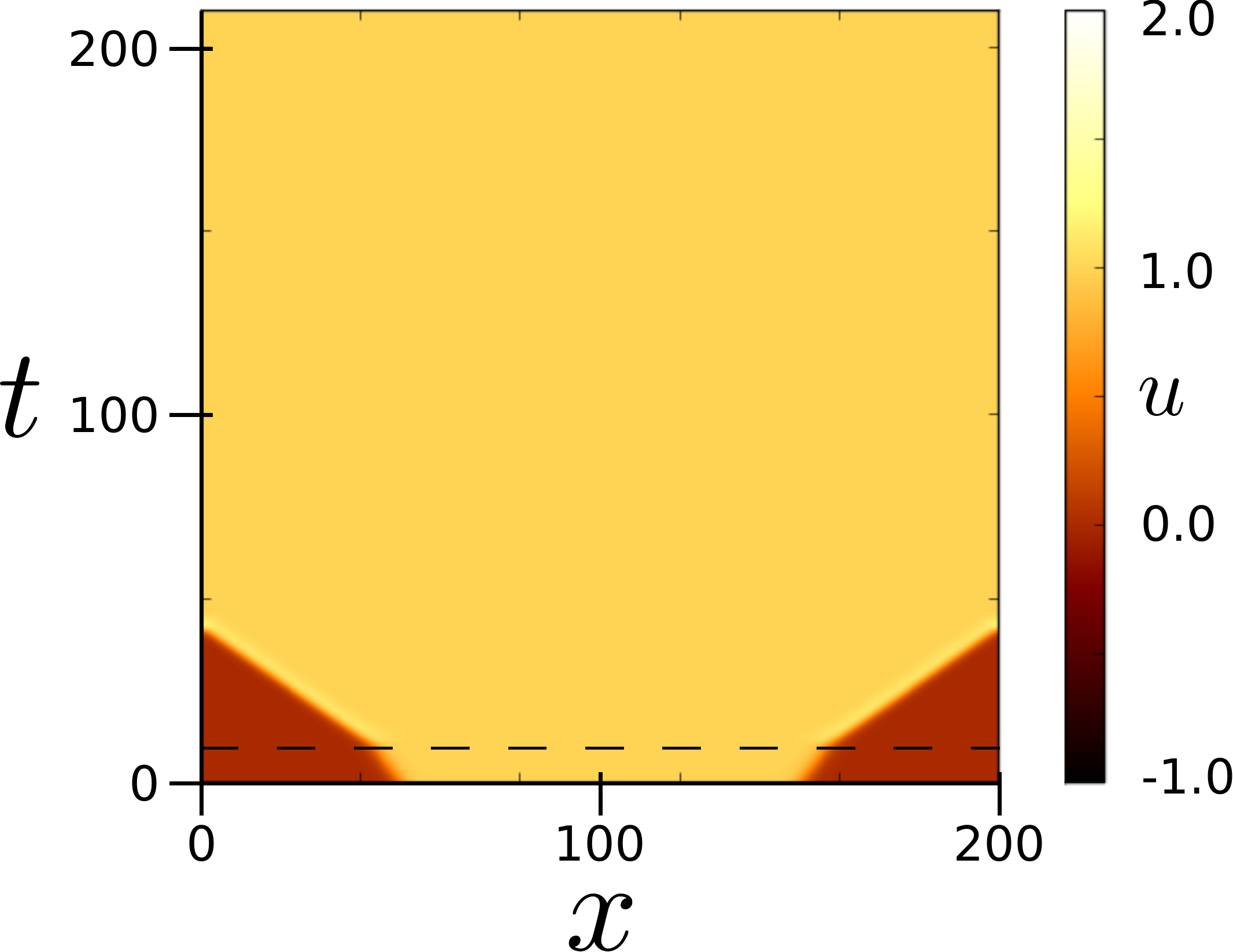,width=5cm}
  }{
    \psfig{file=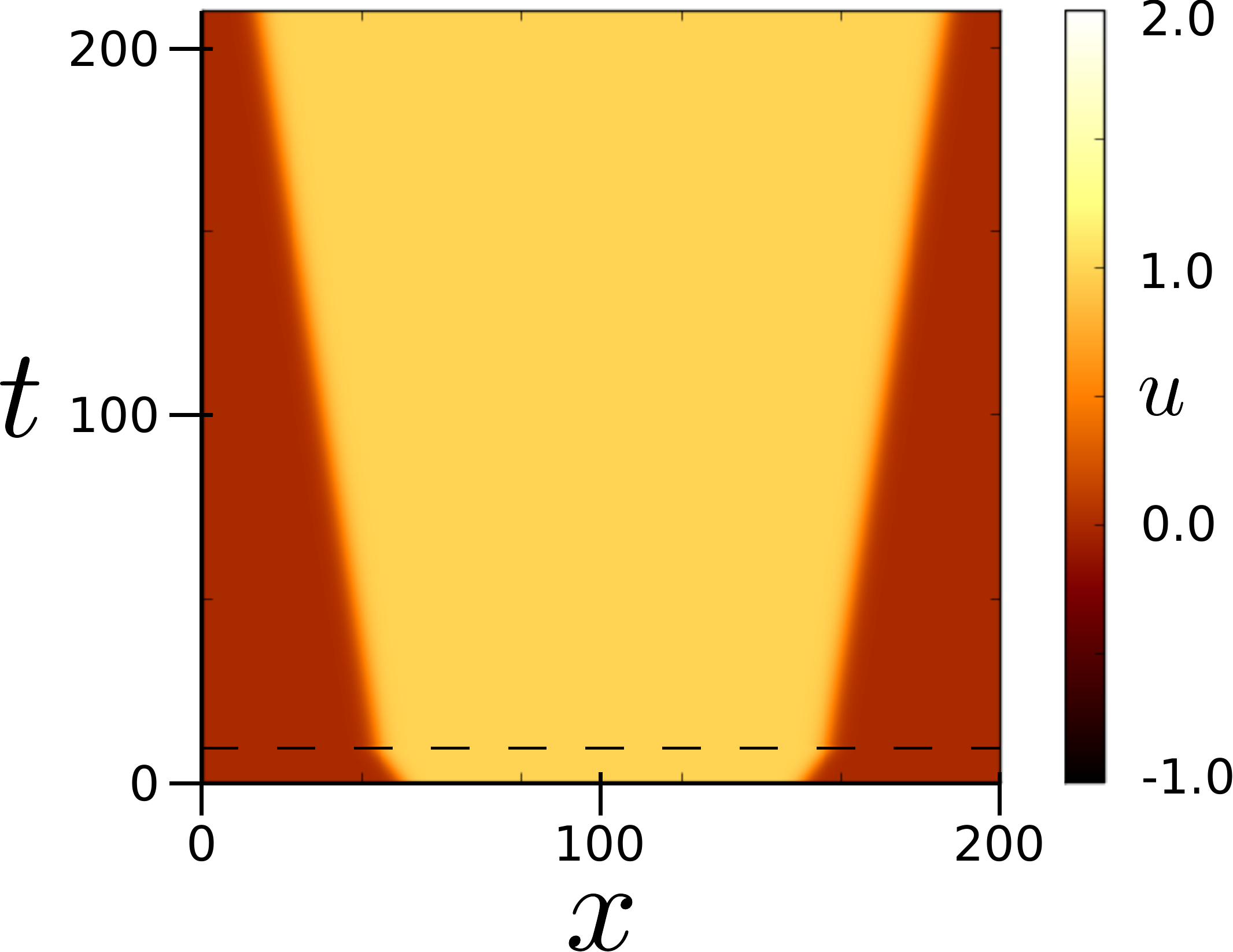,width=5cm}
  }
  \sidebyside{
     (c)\hspace*{5cm}
  }{
     (d)\hspace*{5cm}
  }  
  \sidebyside{
    \psfig{file=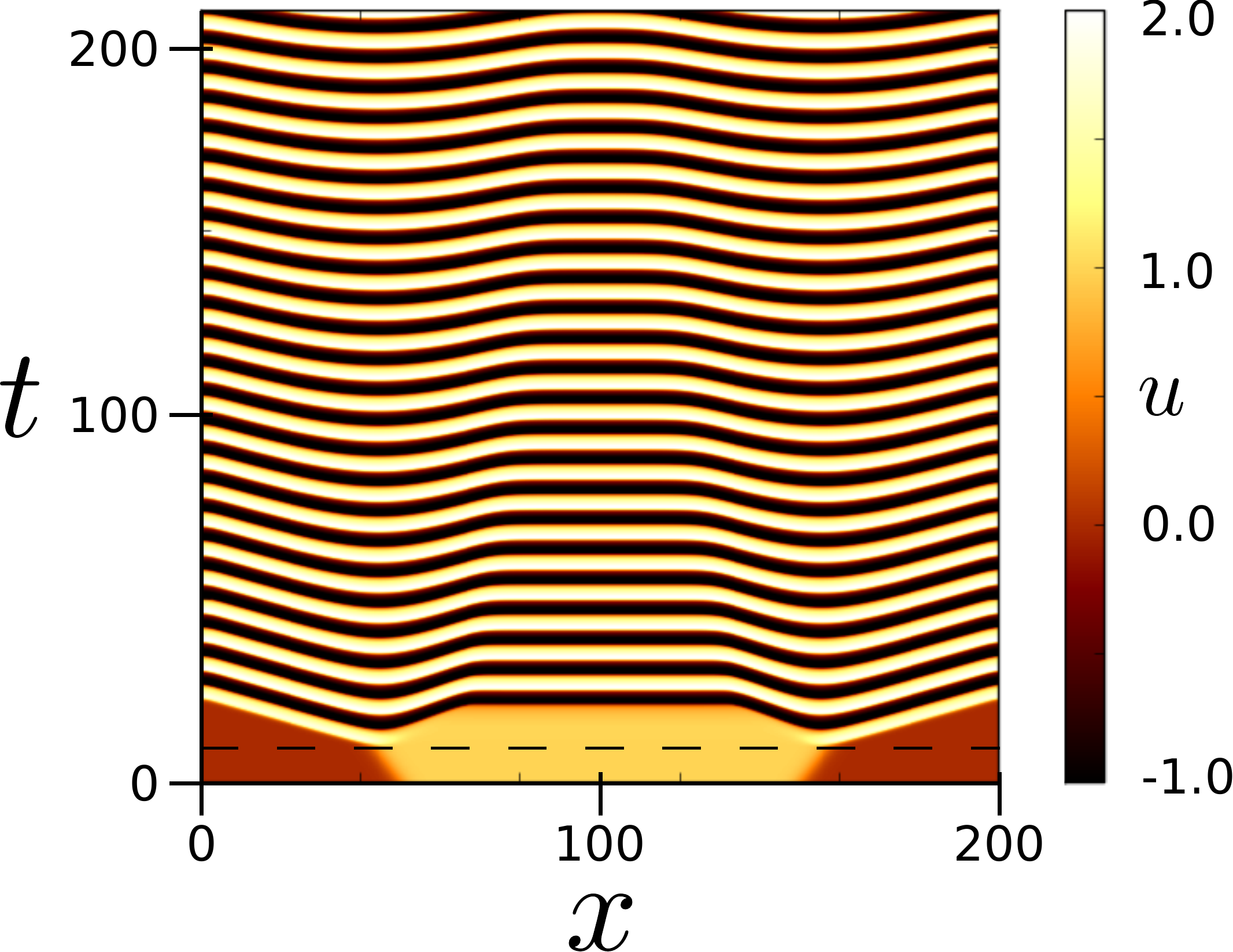,width=5cm}
  }{
    \psfig{file=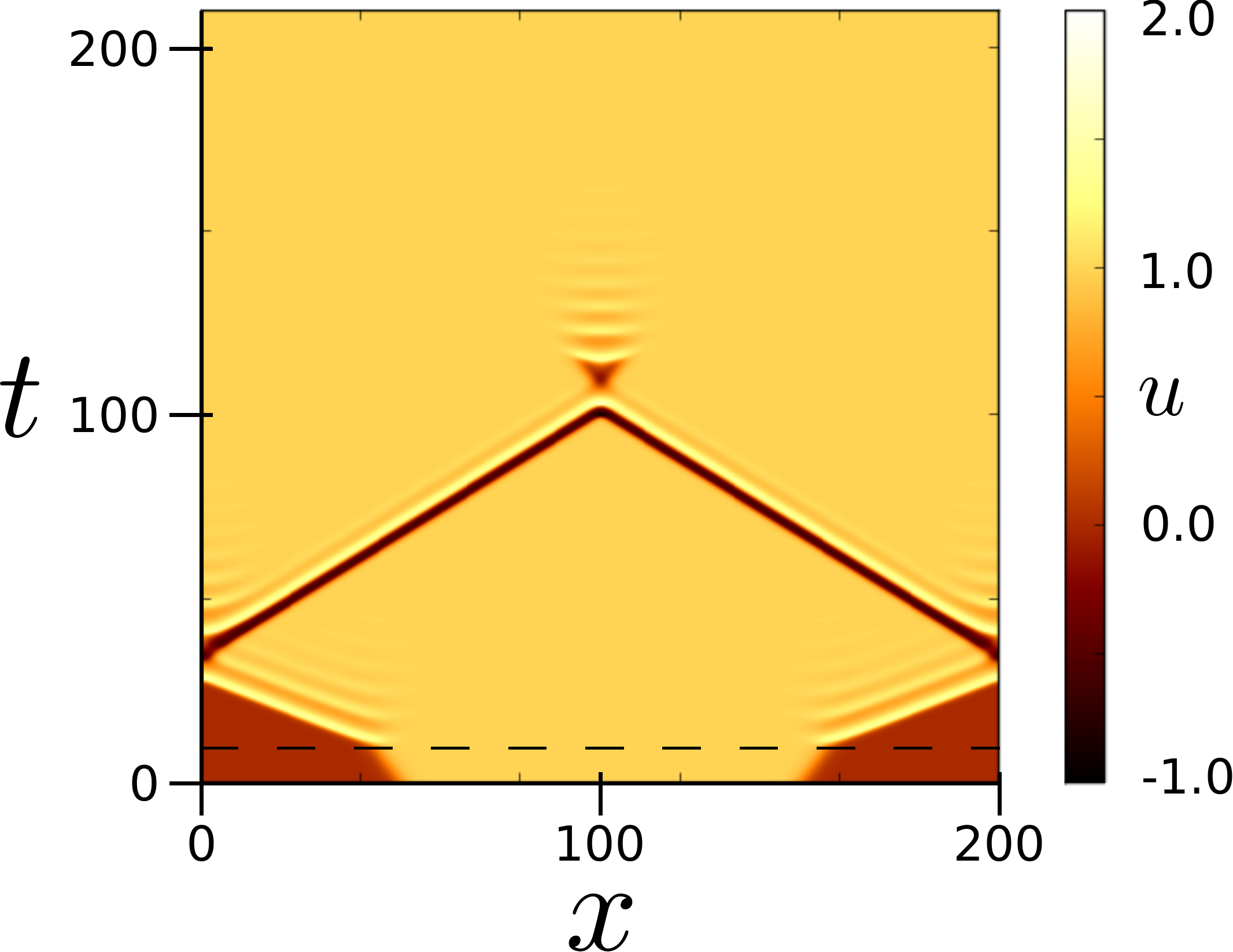,width=5cm}
  }
  \sidebyside{
     (e)\hspace*{5cm}
  }{
     (f)\hspace*{5cm}
  }
  \sidebyside{
    \psfig{file=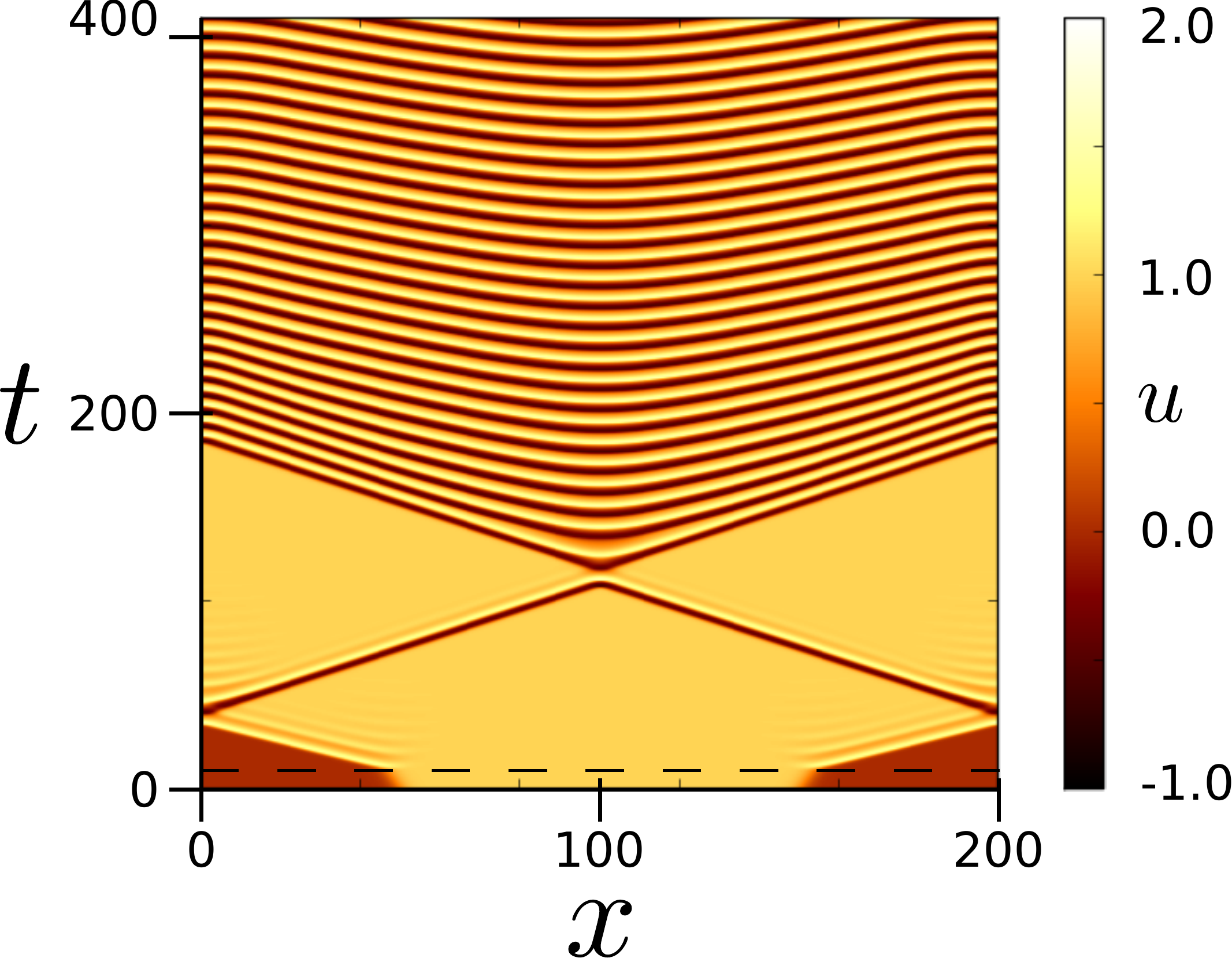,width=5cm}
  }{
    \psfig{file=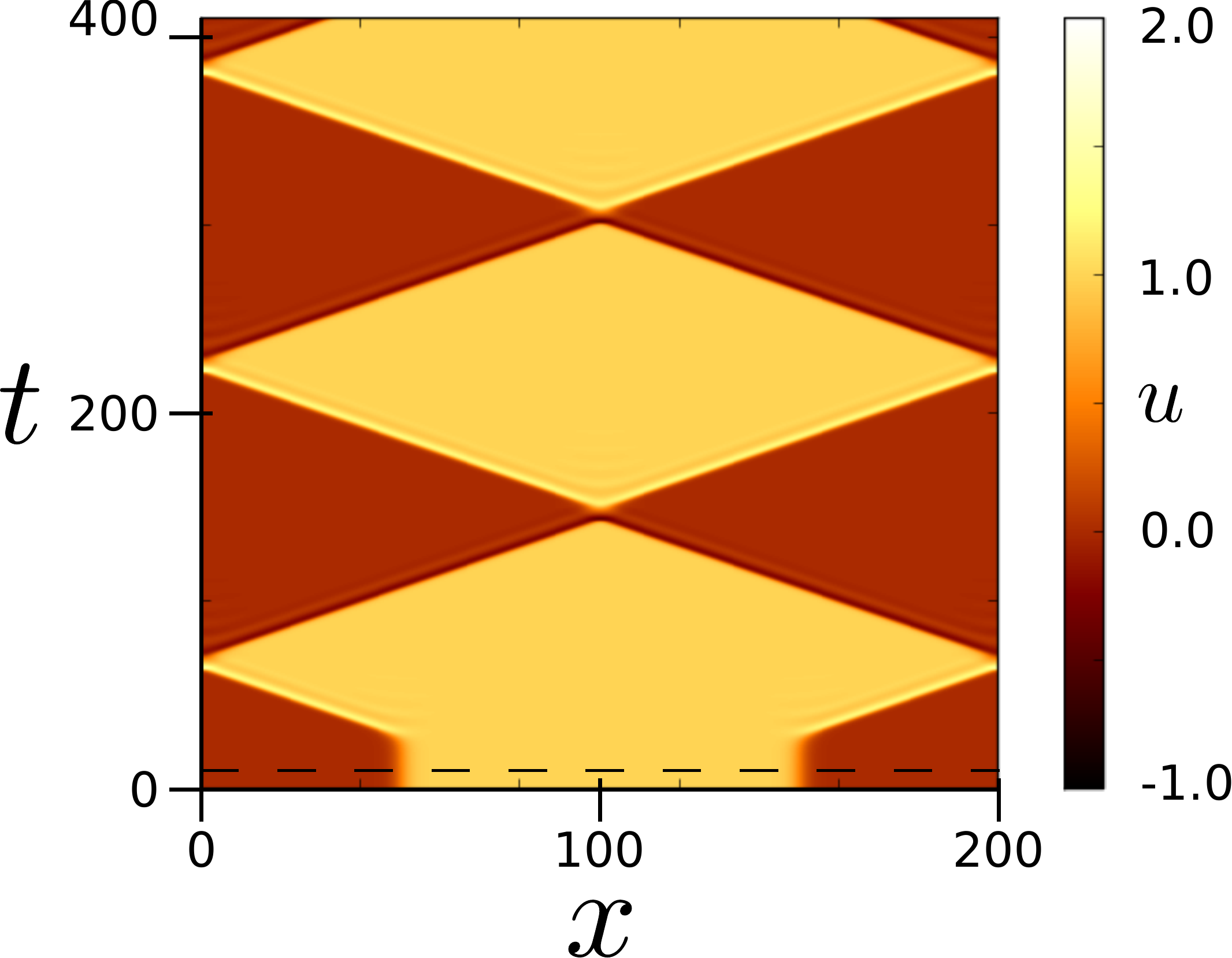,width=5cm}
  }
  \sidebyside{
     (g)\hspace*{5cm}
  }{
     (h)\hspace*{5cm}
  }  
  \sidebyside{
    \psfig{file=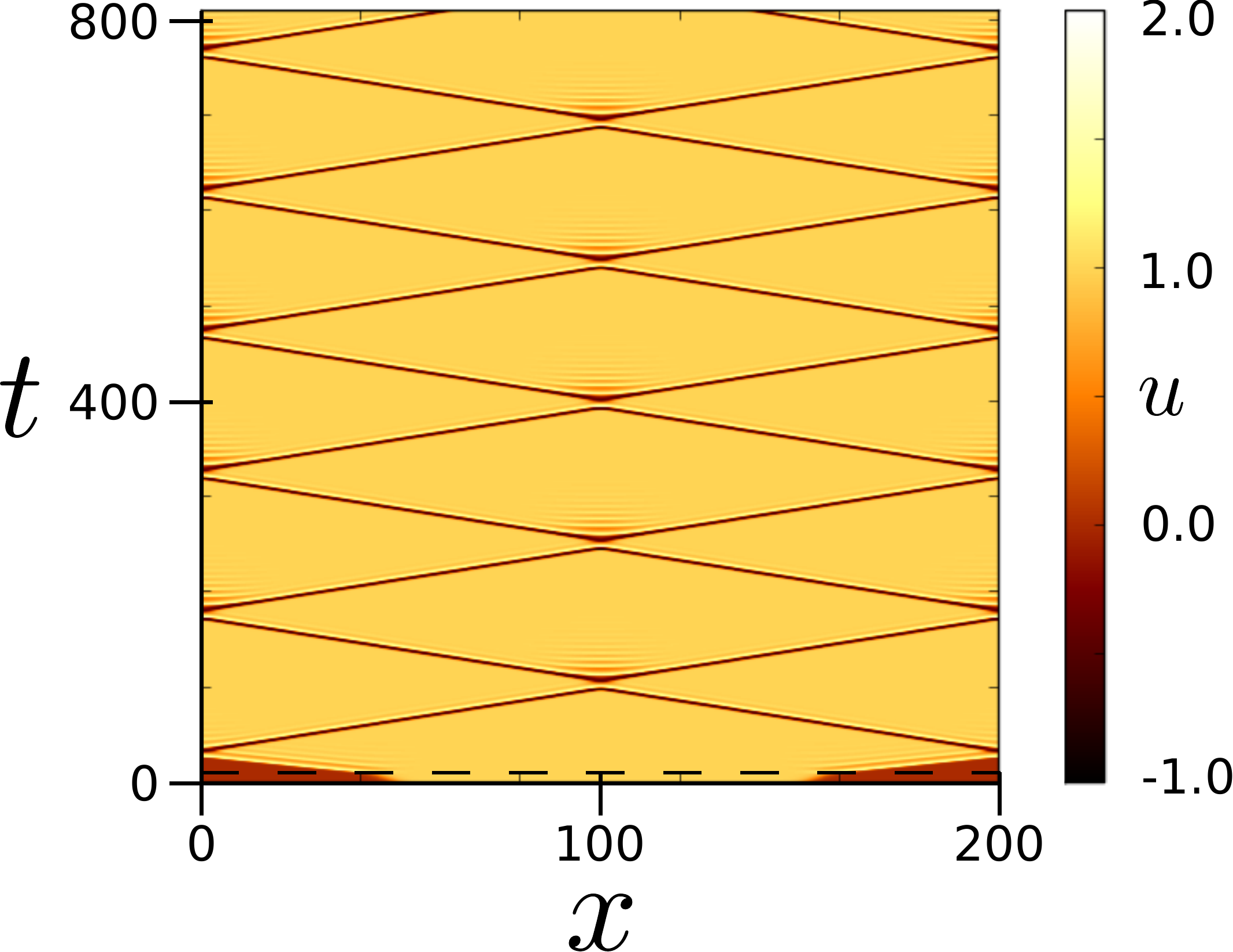,width=5cm}
  }{
    \psfig{file=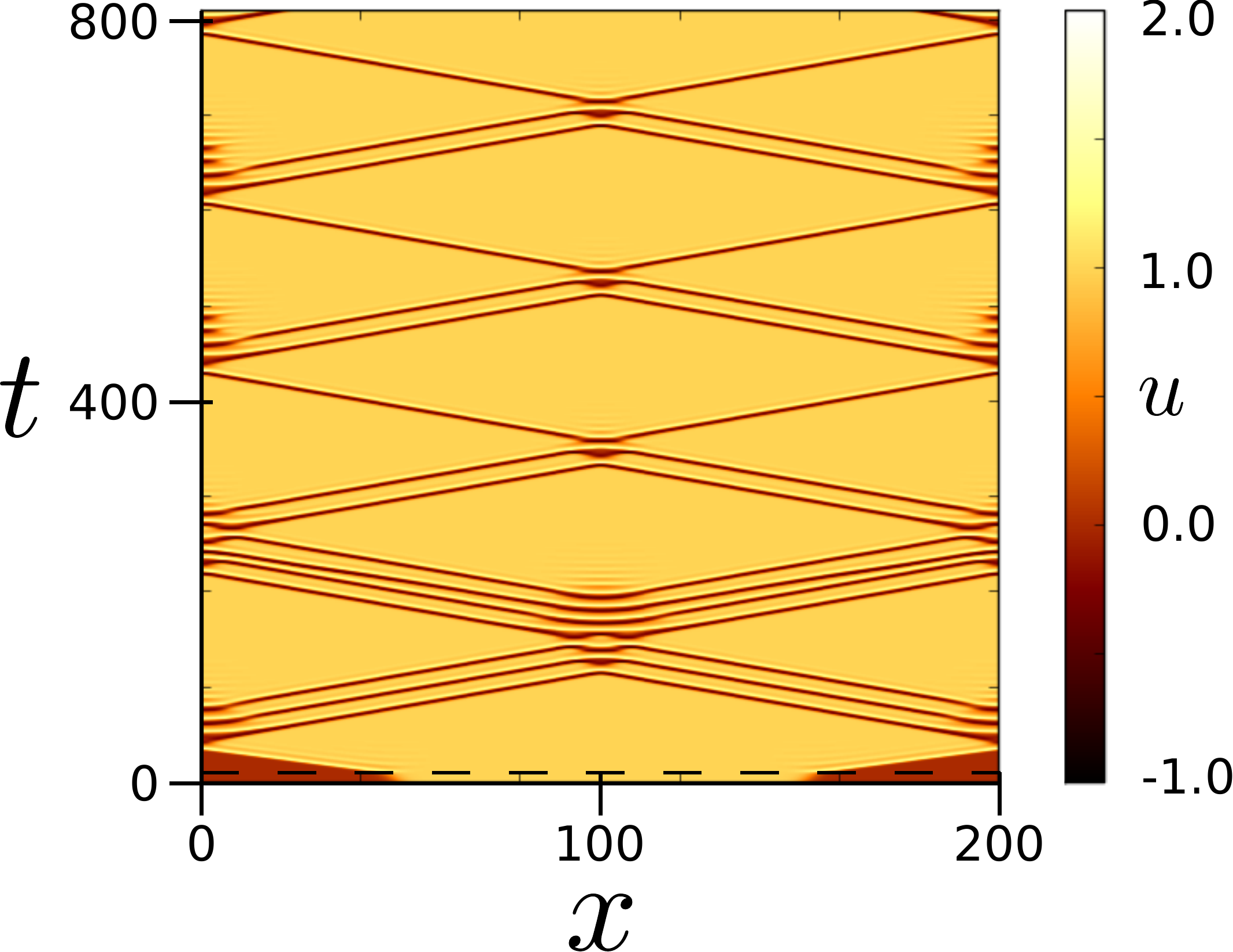,width=5cm}
  }
  
  \caption{Space-time patterns for distributed time-delayed feedback: (a) acceleration of the fronts; (b) deceleration of the fronts; (c) waves; (d) transient traveling pulse leading to homogeneous steady state $u^*_+$; (e) transient traveling pulse leading to waves; (f) pairs of reflected fronts; (g) pairs of traveling pulses; (h) multiple fronts.
  Results obtained via numerical simulation of Eq.\ref{Eq.td} with: (a--c) Strong gamma delay kernel $G(t)= a^2t\exp(-at)$, $a=1$, $\alpha=0$, (a) $\sigma=-0.5$, (b) $\sigma=2$, (c) $\sigma=-2$; (d--h) Weak gamma delay kernel $G(t) = a\exp(-at)$, (d) $a=1$, $\alpha=0$ $\sigma=-1.82$. (e) $a=1$, $\alpha=0.25$ $\sigma=-1.65$, (f) $a=1$, $\alpha=0.5$ $\sigma=-1.4$, (g) $a=1$, $\alpha=0$ $\sigma=-1.84$, (h) $a=1$, $\alpha=0.25$ $\sigma=-1.6$. Time scale (a--d) $0\leq t\leq 210$; (e,f) $0\leq t\leq 410$; (g,h) $0\leq t\leq 810$. Space scale (a--g) $0\leq x\leq 200$. }
  \label{fig.sieb.td}
\end{figure}

(i) The system exhibits two propagating fronts whose respective velocities can be either accelerated or decelerated, see Figs.\ref{fig.sieb.nl}(a,b,c) and \ref{fig.sieb.td}(a,b). This behavior typically arises in the bistability regime in Fig.\ref{fig.sieb.dr.as}. Note 
that for symmetric nonlocal kernels and time-delayed kernels both fronts are equally accelerated or decelerated. On the contrary, asymmetric nonlocal kernels lead to acceleration of one front and deceleration of the other.

(ii) The system exhibits new global spatio-temporal patterns such as Turing patterns, see Fig.\ref{fig.sieb.nl}(d), traveling wave patterns, see Fig.\ref{fig.sieb.nl}(e) or mixed wave patterns, see Fig.\ref{fig.sieb.td}(c). This behavior is characteristic of the wave regime in Fig.\ref{fig.sieb.dr.as}.

(iii) Between these two regimes, the system may exhibit transient patterns before asymptotically approaching a globally stable state. Fig.\ref{fig.sieb.td}(d,e) illustrates these transients for time-delayed feedback. 
The system can also exhibit localized patterns such as coexistence of a homogeneous steady state with Turing or traveling wave patterns Fig.\ref{fig.sieb.nl}(f,g), fronts reflected at the boundaries Fig.\ref{fig.sieb.td}(f), or traveling pulses Fig.\ref{fig.sieb.td}(g,h). All these patterns have been observed in the parameter regime of the dotted area of Fig.\ref{fig.sieb.dr.as}. However, 
these localized patterns cannot be derived from the linear stability analysis of the homogeneous steady state.

Analytical results on the change of the front velocity by the feedback, and a more detailed discussion of the asymmetric
kernels are given elsewhere \cite{SIE14}.

\section{Position control of traveling front solutions to the Schl\"ogl model}\label{sec3}

\begin{figure}[ht]
  \centerline{
    \psfig{file=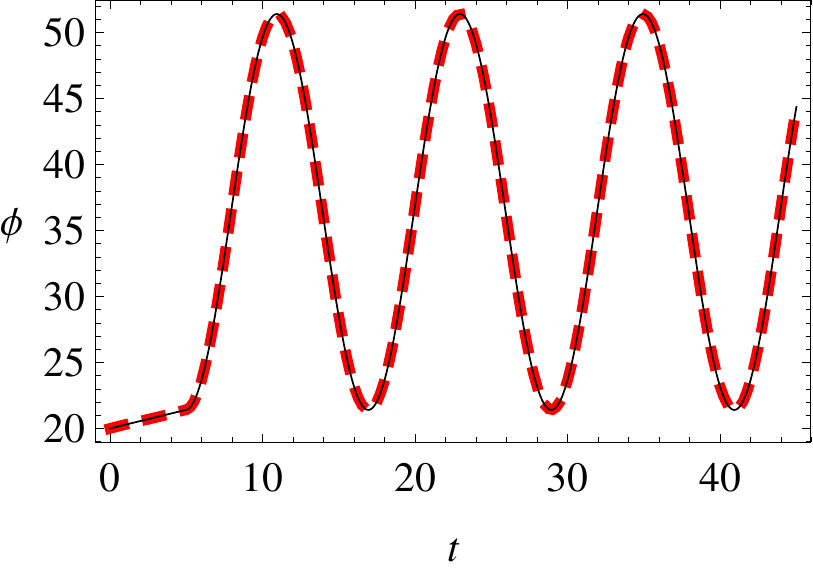,width=6.25cm}
    \psfig{file=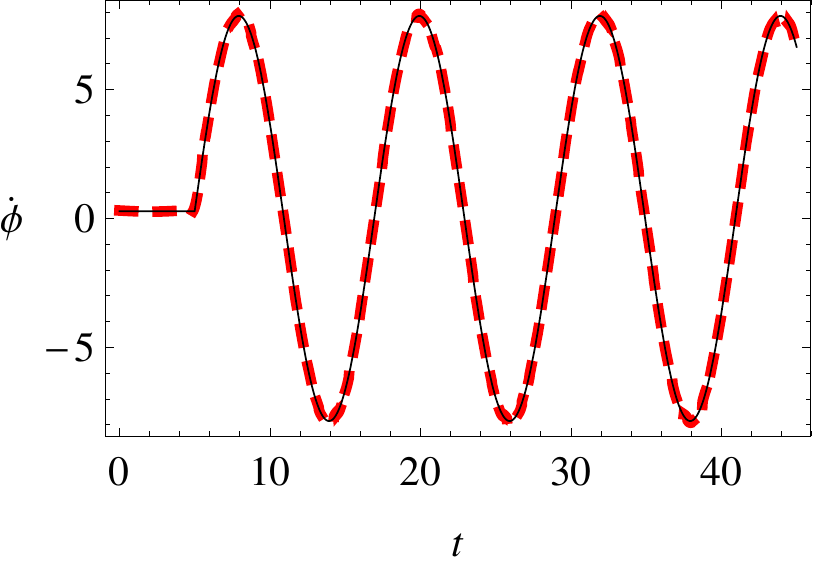,width=6.25cm}
  }
  \caption{Position control of a Schl\"ogl front solution for an additive coupling function $\mathcal{G}=1$. Position (left) and velocity over time data (right)
  obtained from numerical simulations of the controlled 
  Schl\"ogl front (red line) are in excellent agreement with the sinusoidal analytical protocol (black line). The parameter is $\alpha=0.3$.}
  \label{AdditiveControl}
\end{figure}

\begin{figure}[ht]
  \centerline{
    \psfig{file=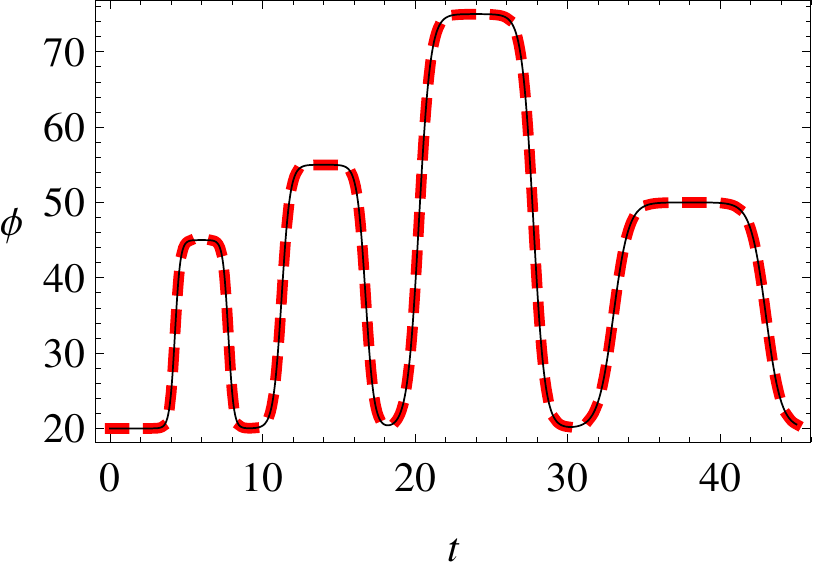,width=6.25cm}
    \psfig{file=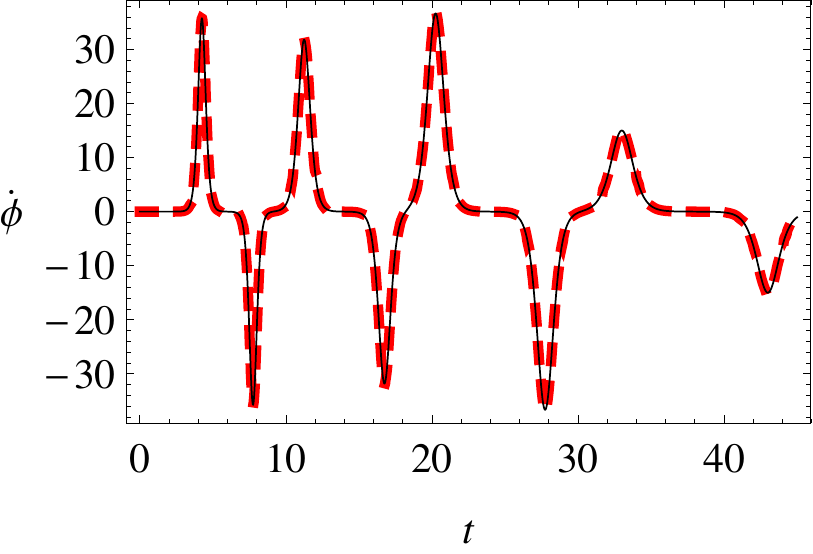,width=6.25cm, height=4.4cm}
  }
  \caption{
  Comparison of the protocol with position (left) and velocity (right) over time data
  obtained by numerical simulation. The control moves a Schl\"ogl front to a certain position, 
  stops it there for a specified time interval, and moves it back (left). A coupling function 
  $\mathcal{G}\left(u\right)=b+du$ with $b=1,\,d=1/2$ motivated by a liquid crystal light valve 
  illuminated by a spatio-temporal light intensity is assumed. The parameter is $\alpha=0.3$.}
  \label{MultiplicativeControl}
\end{figure}

Here we consider the controlled Schl\"ogl model of the form Eq. \eqref{eq:ControlledReactionDiffusionEquation}.
We pursue a perturbative approach to the control problem and interpret the spatio-temporal control function 
$f\left(x,t\right)$ as a small term perturbing a stable traveling front solution $U_c\left(x\right)$.
By multiple scale perturbation theory for small perturbation amplitude $\epsilon$, the following
equation of motion (EOM) for the position $\phi\left(t\right)$ of
the perturbed front can be obtained \cite{schimanskygeier1983effect, engel1985noise, engel1987interaction, 
kulka1995influence, bode1997front, alonso2010wave, loeber2012front}, 
\begin{align}
\dot{\phi} & =c-\frac{\epsilon}{K_{c}}\int_{-\infty}^{\infty}dxe^{cx/D}U_{c}'\left(x\right)\mathcal{G}\left(U_{c}\left(x\right)\right)f\left(x+\phi,t\right),\label{eq:EquationOfMotion}
\end{align}
where $K_{c}=\intop_{-\infty}^{\infty}dxe^{cx/D}\left(U_{c}'\left(x\right)\right)^{2}$ is constant
and $\phi\left(t_{0}\right)=\phi_{0}$ denotes the initial condition. The derivation of Eq. \eqref{eq:EquationOfMotion}
does not define a position of a front a priori, therefore we identify the point of steepest slope with the front position.
\newline
Here, we do not perceive Eq. (\ref{eq:EquationOfMotion}) as an ordinary differential
equation for the position $\phi\left(t\right)$ of the wave under
the given perturbation $f$. Instead, Eq. (\ref{eq:EquationOfMotion})
is viewed as an integral equation for the \textit{control function}
$f$ \cite{Loeber1}. The idea is to find a control which solely drives propagation
in space according to an arbitrary prescribed \textit{protocol of motion}
$\phi\left(t\right)$. Simultaneously, we expect $f$ to prevent large
deformations in the uncontrolled wave profile $U_{c}\left(x\right)$.
We assume that the wave moves unperturbed until reaching position
$\phi_{0}$ at time $t_{0}$, upon which the control is switched on.\\
A general solution of the integral equation Eq. (\ref{eq:EquationOfMotion})
for the control $f$ corresponding to the protocol of motion
$\phi\left(t\right)$ is
\begin{align}
f\left(x,t\right) & =\left(c-\dot{\phi}\right)\dfrac{K_{c}}{G_{c}}\,\mathcal{G}^{-1}\left(U_{c}\left(x-\phi\right)\right)h\left(x-\phi\right),\label{eq:GeneralControl}
\end{align}
with constant $G_{c}=\intop_{-\infty}^{\infty}dxe^{cx/D}U_{c}'\left(x\right)h\left(x\right)$.
Here $\mathcal{G}^{-1}$ denotes the reciprocal of $\mathcal{G}$.
The profile $\mathcal{G}^{-1}h$ of $f$ is co-moving
with the controlled wave and has constant amplitude. 
Eq. (\ref{eq:GeneralControl}) contains
a so far undefined arbitrary function $h\left(x\right)$. A control
proportional to the Goldstone mode $U_{c}'$ shifts the front as a
whole, simultaneously preventing large deformations of the wave profile.
Therefore, in the following we choose $h\left(x\right)=U_{c}'\left(x\right)$,
i.e. 
\begin{align}
f\left(x,t\right) & =\left(c-\dot{\phi}\right)\mathcal{G}^{-1}\left(U_{c}\left(x-\phi\right)\right)U_{c}'\left(x-\phi\right).\label{eq:SpatiotemporalControl}
\end{align}
The constant $K_{c}/G_{c}$ cancels out because $K_{c}=G_{c}$ for this choice. The control amplitude is entirely determined by the time dependent coefficient $c-\dot{\phi}$. Note that only the coupling function $\mathcal{G}$, the velocity $c$ of the uncontrolled front,
and the derivative of traveling wave profile $U_c$ enters the expression for the control signal. In particular, no knowledge of the underlying reaction kinetics $R\left(u\right)$ or parameter values is 
necessary. This makes the method useful for applications where model equations are only approximately known but the wave profile can be measured with sufficient accuracy.
\newline
For a first sanity check, we assume a spatio-temporal control of the parameter $\alpha$ of the rescaled Schl\"ogl model Eq. \eqref{eq:RescaledSchloeglTravelingFrontProfile}.
This parameter can be seen as a measure for the excitation threshold of the Schl\"ogl model: 
a localized perturbation nowhere exceeding this value cannot trigger a transition from the lower stationary stable state $u=0$ to the 
upper stationary stable state $u=1$. Substituting $\alpha\rightarrow \alpha+ f\left(x,t\right)$
in Eq. \eqref{eq:RescaledReactionFunction} yields a coupling function
\begin{align}
\mathcal{G}\left(u\right) & = u\left(u-1\right).
\end{align}
The solution Eq. \eqref{eq:SpatiotemporalControl} for the control function immediately leads to
\begin{align}
f\left(x,t\right) & =\left(c-\dot{\phi}\right)\frac{U_{c}'\left(x-\phi\right)}{U_{c}\left(x-\phi\right)\left[U_{c}\left(x-\phi\right)-1\right]}.
\end{align}
Plugging in the traveling front solution Eq. \eqref{eq:RescaledSchloeglTravelingFrontProfile}, we find a space-independent control function,
\begin{align}
f\left(x,t\right) & =\frac{1}{\sqrt{2}}\left(c-\dot{\phi}\right). \label{eq:Constanta}
\end{align}
We interpret the result as follows. We set out to find a spatio-temporal control which changes the velocity of a traveling wave in a prescribed way
and simultaneously preserves the uncontrolled front profile $U_c$. Because the front velocity $c=\frac{1}{\sqrt{2}}\left(1-2\alpha\right)$ depends linearly on $\alpha$ with 
coefficient $-\sqrt{2}$, we expect that an increase of parameter $\alpha$ by $\Delta \alpha=- \frac{1}{\sqrt{2}}\Delta c$ to a value $\tilde{\alpha} = \alpha +\Delta \alpha$ 
yields a front velocity $\tilde{c} = c + \Delta c$. Indeed, substituting $\phi\left(t\right) =  \tilde{c} t = \left(c + \Delta c\right) t$ in Eq. \eqref{eq:Constanta} 
yields $f\left(x,t\right) =-\frac{\Delta c}{\sqrt{2}}=\Delta \alpha$. Because in the Schl\"ogl model, the front profile does not depend on the parameter $\alpha$, we achieved the 
desired goal of changing the front velocity while preserving the uncontrolled front profile for the case of constant protocol velocities $\dot{\phi}=\tilde{c}=\text{const.}$
The result Eq. \eqref{eq:Constanta} can be seen as a generalization to 
arbitrary protocols with non-constant protocol velocities. As long as changes in the protocol velocity are slow, $|\ddot{\phi}| \ll 1$, and the 
maximum and minimum protocol velocities are sufficiently close to the velocity $c$ of the uncontrolled front, we expect
Eq. \eqref{eq:Constanta} to result in a successful position control. Both assumption of slow changes in the front velocity and sufficiently small amplitude of perturbations 
are inherent in the multiple scale perturbation approach leading to the EOM \eqref{eq:EquationOfMotion} for perturbed traveling waves.
\newline
To demonstrate the performance of the proposed control approach, we consider an additive control $\mathcal{G}\left(u\right)=1$ with a sinusoidal protocol
\begin{align}
\phi\left(t\right) & =B_0 +A\sin\left( 2\pi t/T + B_1\right).\label{eq:SinusoidalProtocol}
\end{align}
$B_0$ and $B_1$ are determined by $\phi\left(t_0\right) = \phi_0,\,\dot{\phi}\left(t_0\right) = c$ such that the protocol is smooth at the initial time $t_0$. 
\newline
We carried out numerical simulations of the controlled Schl\"ogl front with
no-flux boundary conditions and using the uncontrolled wave profile $U_c\left(x\right)$ as the initial condition.
In Fig. \ref{AdditiveControl}, the obtained position and velocity over time data are compared with the 
prescribed protocol $\phi \left(t\right)$.
The controlled front follows the prescribed protocol very closely. The front profile is only slightly deformed 
by the control and thus the front profile of the uncontrolled traveling wave is preserved. Furthermore, we showed by examples that the control Eq. \eqref{eq:SpatiotemporalControl} 
is close to a numerically computed optimal control \cite{Loeber1} with a traveling wave solution $U_c\left(x-\phi\left(t\right)\right)$ shifted according to the protocol $\phi$ as 
the desired distribution enforced on the reaction-diffusion system \cite{buchholz2013on}.
\newline
As a second example, we consider a coupling function motivated by the controlled liquid crystal light valve, Eq. \eqref{eq:LCLVBistableSystem},
$\mathcal{G}\left(u\right)=b+du$.
We define a smooth step function
\begin{align}
\Theta_k\left(t\right) & =\left(1+\tanh\left( k t\right)\right)/2.\label{eq:SmoothStep}
\end{align}
The usual discontinuous step function is recovered in the limit $\lim_{k\rightarrow\infty}\Theta_k\left(t\right)$.
A smooth box function is defined as
\begin{align}
B_k\left(t\right) & =\Theta_k\left(1/2-t\right)+\Theta_k\left(t+1/2\right)-1.\label{eq:SmoothBox}
\end{align}
Using Eqs. \eqref{eq:SmoothStep} and \eqref{eq:SmoothBox}, we specify a protocol which moves the front to a certain position $\phi_0 + A_i$, stops it there for a given time interval $w_i$, moves it back
to the initial position $\phi_0$ and so on. The protocol consists of boxes of width $w_i$ and amplitude $A_i$ at times $t_i$,
\begin{align}
\phi\left(t\right) & =\phi_0 + \sum_i A_i B_k\left((t-t_i)/w_i\right).\label{eq:BoxProtocol}
\end{align}
The front follows the prescribed protocol (black line) very closely, as the comparison with numerically obtained (red line) position and velocity (right) over time data shows, see Fig. \ref{MultiplicativeControl}.

\section{Conclusions}\label{sec4}

Using the Schl\"ogl model as a paradigmatic example of a bistable reaction-diffusion system, we have discussed some physically feasible options of open and closed loop spatio-temporal control of RD systems. Control constraints arise from the physical meaning of the involved state and control variables representing concentrations, 
temperature, etc. Certainly, this aspect deserves more attention and further investigation with regard to the development of easy-to-realize experimental control schemes for RD systems.
\newline
We have shown by two different control approaches, closed-loop and open-loop control, how position and velocity of a chemical front can be precisely controlled in space and time without deforming the front profile substantially, and how a variety of spatio-temporal patterns can be generated by feedback control. Our open-loop approach applies, if the front profile can be measured in the uncontrolled RD system with a precision that allows one to determine its spatial derivative with sufficient accuracy, Eq. \eqref{eq:SpatiotemporalControl}. In addition, the control function $\mathcal{G}$ that 
depends on the intended control loop must be known; as far as possible zeros in $\mathcal{G}$ should be avoided. Remarkably, the often incompletely known 
reaction rate is not required in order to set up the control. This makes the approach very promising if the underlying RD dynamics cannot be revealed in 
all details. Moreover, in two spatial dimensions, a generalized version of our approach allows for a precise and efficient control of the shape of wave patterns too. In case of the closed-loop control special effects arise for asymmetric nonlocal kernels \cite{SIE14}.

\bibliographystyle{ws-book-har}    	
\bibliography{literature}         

\begin{thebibliography}{64}
\newcommand{\enquote}[1]{#1}
\providecommand{\natexlab}[1]{#1}
\providecommand{\url}[1]{\texttt{#1}}
\providecommand{\urlprefix}{URL }
\expandafter\ifx\csname urlstyle\endcsname\relax
  \providecommand{\doi}[1]{doi:\discretionary{}{}{}#1}\else
  \providecommand{\doi}{doi:\discretionary{}{}{}\begingroup
  \urlstyle{rm}\Url}\fi

\bibitem[{Agladze \emph{et~al.}(1987)Agladze, Davydov and
  Mikhailov}]{agladze1987observation}
Agladze, K., Davydov, V. and Mikhailov, A. (1987). \enquote{An observation of
  resonance of spiral waves in distributed excitable medium,} \emph{JETP Lett.}
  \textbf{45}, 12, pp. 767--770.

\bibitem[{Alonso \emph{et~al.}(2010)Alonso, L{\"o}ber, B{\"a}r and
  Engel}]{alonso2010wave}
Alonso, S., L{\"o}ber, J., B{\"a}r, M. and Engel, H. (2010). \enquote{Wave
  propagation in heterogeneous bistable and excitable media,} \emph{Europ.
  Phys. J. ST} \textbf{187}, 1, pp. 31--40.

\bibitem[{B{\'a}ns{\'a}gi \emph{et~al.}(2011)B{\'a}ns{\'a}gi, Vanag and
  Epstein}]{bansagi2011tomography}
B{\'a}ns{\'a}gi, T., Vanag, V.~K. and Epstein, I.~R. (2011).
  \enquote{Tomography of reaction-diffusion microemulsions reveals
  three-dimensional turing patterns,} \emph{Science} \textbf{331}, 6022, pp.
  1309--1312.

\bibitem[{B{\"a}r \emph{et~al.}(1992)B{\"a}r, Falcke, Z{\"u}licke, Engel,
  Eiswirth and Ertl}]{bar1992reaction}
B{\"a}r, M., Falcke, M., Z{\"u}licke, C., Engel, H., Eiswirth, M. and Ertl, G.
  (1992). \enquote{Reaction fronts and pulses in the {CO} oxidation on {Pt}:
  theoretical analysis,} \emph{Surf. Sci.} \textbf{269}, pp. 471--475.

\bibitem[{Bode(1997)}]{bode1997front}
Bode, M. (1997). \enquote{Front-bifurcations in reaction-diffusion systems with
  inhomogeneous parameter distributions,} \emph{Physica D} \textbf{106}, 3, pp.
  270--286.

\bibitem[{Buchholz \emph{et~al.}(2013)Buchholz, Engel, Kammann and
  Tr\"{o}ltzsch}]{buchholz2013on}
Buchholz, R., Engel, H., Kammann, E. and Tr\"{o}ltzsch, F. (2013). \enquote{On
  the optimal control of the {S}chl\"{o}gl-model,} \emph{Comput. Optim. Appl.}
  \textbf{56}, pp. 153--185.

\bibitem[{Colet \emph{et~al.}(2014)Colet, Matias, Gelens and Gomila}]{COL13}
Colet, P., Matias, M.~A., Gelens, L. and Gomila, D. (2014). \enquote{Formation
  of localized structures in bistable systems through nonlocal spatial
  coupling. {I}. {G}eneral framework,} \emph{Phys. Rev. E} \textbf{89}, p.
  012914.

\bibitem[{Dahlem \emph{et~al.}(2008)Dahlem, Schneider and Sch{\"o}ll}]{DAH08}
Dahlem, M.~A., Schneider, F.~M. and Sch{\"o}ll, E. (2008). \enquote{{Failure of
  feedback as a putative common mechanism of spreading depolarizations in
  migraine and stroke},} \emph{Chaos} \textbf{18}, p. 026110.

\bibitem[{Engel(1985)}]{engel1985noise}
Engel, A. (1985). \enquote{Noise-induced front propagation in a bistable
  system,} \emph{Phys. Lett. A} \textbf{113}, 3, pp. 139--142.

\bibitem[{Engel and Ebeling(1987)}]{engel1987interaction}
Engel, A. and Ebeling, W. (1987). \enquote{Interaction of moving interfaces
  with obstacles,} \emph{Phys. Lett. A} \textbf{122}, 1, pp. 20--24.

\bibitem[{{\'E}rdi(1989)}]{erdi1989mathematical}
{\'E}rdi, P. (1989). \emph{Mathematical models of chemical reactions: theory
  and applications of deterministic and stochastic models} (Manchester
  University Press, Manchester).

\bibitem[{Gelens \emph{et~al.}(2014)Gelens, Matias, Gomila, Dorissen and
  Colet}]{COL14}
Gelens, L., Matias, M.~A., Gomila, D., Dorissen, T. and Colet, P. (2014).
  \enquote{Formation of localized structures in bistable systems through
  nonlocal spatial coupling. {II}. {T}he nonlocal {G}inzburg-{L}andau
  equation,} \emph{Phys. Rev. E} \textbf{89}, p. 012915.

\bibitem[{Hanna \emph{et~al.}(1982)Hanna, Saul and
  Showalter}]{hanna1982detailed}
Hanna, A., Saul, A. and Showalter, K. (1982). \enquote{Detailed studies of
  propagating fronts in the iodate oxidation of arsenous acid,} \emph{J. Am.
  Chem. Soc.} \textbf{104}, 14, pp. 3838--3844.

\bibitem[{Haudin \emph{et~al.}(2009)Haudin, Elias, Rojas, Bortolozzo, Clerc and
  Residori}]{haudin2009driven}
Haudin, F., Elias, R., Rojas, R., Bortolozzo, U., Clerc, M. and Residori, S.
  (2009). \enquote{Driven front propagation in 1d spatially periodic media,}
  \emph{Phys. Rev. Lett.} \textbf{103}, 12, p. 128003.

\bibitem[{Haudin \emph{et~al.}(2010)Haudin, El{\'\i}as, Rojas, Bortolozzo,
  Clerc and Residori}]{haudin2010front}
Haudin, F., El{\'\i}as, R., Rojas, R., Bortolozzo, U., Clerc, M. and Residori,
  S. (2010). \enquote{Front dynamics and pinning-depinning phenomenon in
  spatially periodic media,} \emph{Phys. Rev. E} \textbf{81}, 5, p. 056203.

\bibitem[{Kang \emph{et~al.}(2002)Kang, Shelley and Sompolinsky}]{KAN02}
Kang, K., Shelley, M. and Sompolinsky, H. (2002). \enquote{Mexican hats and
  pinwheels in visual cortex,} \emph{PNAS} \textbf{100}, 5, pp. 2848--2853.

\bibitem[{Kapral and Showalter(1995)}]{KAP95a}
Kapral, R. and Showalter, K. (eds.) (1995). \emph{Chemical Waves and Patterns}
  (Kluwer, Dordrecht).

\bibitem[{Kehrt \emph{et~al.}(2009)Kehrt, H{\"o}vel, Flunkert, Dahlem, Rodin
  and Sch{\"o}ll}]{KEH09}
Kehrt, M., H{\"o}vel, P., Flunkert, V., Dahlem, M.~A., Rodin, P. and
  Sch{\"o}ll, E. (2009). \enquote{Stabilization of complex spatio-temporal
  dynamics near a subcritical {H}opf bifurcation by time-delayed feedback,}
  \emph{Eur. Phys. J. B} \textbf{68}, pp. 557--565.

\bibitem[{Kevrekidis \emph{et~al.}(2004)Kevrekidis, Kevrekidis, Malomed,
  Nistazakis and Frantzeskakis}]{kevrekidis2004dragging}
Kevrekidis, P., Kevrekidis, I., Malomed, B., Nistazakis, H. and Frantzeskakis,
  D. (2004). \enquote{Dragging bistable fronts,} \emph{Phys. Scr.} \textbf{69},
  6, p. 451.

\bibitem[{Kim \emph{et~al.}(2001)Kim, Bertram, Pollmann, von Oertzen,
  Mikhailov, Rotermund and Ertl}]{KIM01}
Kim, M., Bertram, M., Pollmann, M., von Oertzen, A., Mikhailov, A.~S.,
  Rotermund, H.~H. and Ertl, G. (2001). \enquote{Controlling chemical
  turbulence by global delayed feedback: Pattern formation in catalytic {CO}
  oxidation on {Pt(110)},} \emph{Science} \textbf{292}, p. 1357.

\bibitem[{Korzukhin(1967{\natexlab{a}})}]{Korzukhin1967mathematicala}
Korzukhin, M. (1967{\natexlab{a}}). \enquote{Mathematical modeling of the
  kinetics of homogeneous chemical systems. {II},} in G.~Frank (ed.),
  \emph{Oscillatory Processes in Biological and Chemical Systems} (Nauk,
  Moscow), pp. 231--242.

\bibitem[{Korzukhin(1967{\natexlab{b}})}]{Korzukhin1967mathematicalb}
Korzukhin, M. (1967{\natexlab{b}}). \enquote{Mathematical modeling of the
  kinetics of homogeneous chemical systems. {III},} in G.~Frank (ed.),
  \emph{Oscillatory Processes in Biological and Chemical Systems} (Nauk,
  Moscow), pp. 242--251.

\bibitem[{Kulka \emph{et~al.}(1995)Kulka, Bode and
  Purwins}]{kulka1995influence}
Kulka, A., Bode, M. and Purwins, H. (1995). \enquote{On the influence of
  inhomogeneities in a reaction-diffusion system,} \emph{Phys. Lett. A}
  \textbf{203}, 1, pp. 33--39.

\bibitem[{Kuramoto(1984)}]{KUR84}
Kuramoto, Y. (1984). \emph{Chemical Oscillations, Waves and Turbulence}
  (Springer-Verlag, Berlin).

\bibitem[{Kyrychko \emph{et~al.}(2009)Kyrychko, Blyuss, Hogan and
  Sch{\"o}ll}]{KYR09}
Kyrychko, Y.~N., Blyuss, K.~B., Hogan, S.~J. and Sch{\"o}ll, E. (2009).
  \enquote{Control of spatio-temporal patterns in the {Gray-Scott} model,}
  \emph{Chaos} \textbf{19}, 4, p. 043126.

\bibitem[{Lebiedz and Brandt-Pollmann(2003)}]{PhysRevLett.91.208301}
Lebiedz, D. and Brandt-Pollmann, U. (2003). \enquote{Manipulation of
  self-aggregation patterns and waves in a reaction-diffusion system by optimal
  boundary control strategies,} \emph{Phys. Rev. Lett.} \textbf{91}, p. 208301.

\bibitem[{L\"ober \emph{et~al.}(2012)L\"ober, B\"ar and
  Engel}]{loeber2012front}
L\"ober, J., B\"ar, M. and Engel, H. (2012). \enquote{Front propagation in
  one-dimensional spatially periodic bistable media,} \emph{Phys. Rev. E}
  \textbf{86}, p. 066210.

\bibitem[{L\"ober and Engel(2014)}]{Loeber1}
L\"ober, J. and Engel, H. (2014). \enquote{Controlling the position of
  traveling waves in reaction-diffusion systems,} arXiv:1304.2327,
  \urlprefix\url{http://arxiv.org/abs/1304.2327}, accepted at Phys. Rev. Lett.

\bibitem[{Malomed \emph{et~al.}(2002)Malomed, Frantzeskakis, Nistazakis,
  Yannacopoulos and Kevrekidis}]{malomed2002pulled}
Malomed, B., Frantzeskakis, D., Nistazakis, H., Yannacopoulos, A. and
  Kevrekidis, P. (2002). \enquote{Pulled fronts in the cahn--hilliard
  equation,} \emph{Phys. Lett. A} \textbf{295}, 5, pp. 267--272.

\bibitem[{Meixner \emph{et~al.}(2000)Meixner, Rodin, Sch{\"o}ll and
  Wacker}]{MEI00b}
Meixner, M., Rodin, P., Sch{\"o}ll, E. and Wacker, A. (2000). \enquote{Lateral
  current density fronts in globally coupled bistable semiconductors with {S}-
  or {Z}-shaped current voltage characteristic,} \emph{Eur.~Phys.~J.~B}
  \textbf{13}, p. 157.

\bibitem[{Mihaliuk \emph{et~al.}(2002)Mihaliuk, Sakurai, Chirila and
  Showalter}]{mihaliuk2002feedback}
Mihaliuk, E., Sakurai, T., Chirila, F. and Showalter, K. (2002).
  \enquote{Feedback stabilization of unstable propagating waves,} \emph{Phys.
  Rev. E} \textbf{65}, 6; PART 1, pp. 065602--65602.

\bibitem[{Mikhailov(1990)}]{mikhailov1990foundations}
Mikhailov, A. (1990). \emph{Foundations of synergetics I: Distributed active
  systems} (Springer-Verlag, New York).

\bibitem[{Mikhailov and Showalter(2006)}]{mikhailov2006control}
Mikhailov, A. and Showalter, K. (2006). \enquote{Control of waves, patterns and
  turbulence in chemical systems,} \emph{Phys. Rep.} \textbf{425}, 2, pp.
  79--194.

\bibitem[{Nettesheim \emph{et~al.}(1993)Nettesheim, Von~Oertzen, Rotermund and
  Ertl}]{nettesheim1993reaction}
Nettesheim, S., Von~Oertzen, A., Rotermund, H. and Ertl, G. (1993).
  \enquote{Reaction diffusion patterns in the catalytic co-oxidation on pt
  (110): Front propagation and spiral waves,} \emph{J. Chem. Phys.}
  \textbf{98}, 12, pp. 9977--9985.

\bibitem[{Nicola \emph{et~al.}(2002)Nicola, Or-Guil, Wolf and B{\"a}r}]{NIC02}
Nicola, E.~M., Or-Guil, M., Wolf, W. and B{\"a}r, M. (2002). \enquote{{Drifting
  pattern domains in a reaction-diffusion system with nonlocal coupling},}
  \emph{Phys. Rev. E} \textbf{65}.

\bibitem[{Nistazakis \emph{et~al.}(2002)Nistazakis, Kevrekidis, Malomed,
  Frantzeskakis and Bishop}]{nistazakis2002targeted}
Nistazakis, H., Kevrekidis, P., Malomed, B., Frantzeskakis, D. and Bishop, A.
  (2002). \enquote{Targeted transfer of solitons in continua and lattices,}
  \emph{Phys. Rev. E} \textbf{66}, 1, p. 015601.

\bibitem[{Papsin \emph{et~al.}(1981)Papsin, Hanna and
  Showalter}]{papsin1981bistability}
Papsin, G.~A., Hanna, A. and Showalter, K. (1981). \enquote{Bistability in the
  iodate oxidation of arsenous acid,} \emph{J. Phys. Chem.} \textbf{85}, 17,
  pp. 2575--2582.

\bibitem[{Plenge \emph{et~al.}(2001)Plenge, Rodin, Sch{\"o}ll and
  Krischer}]{PLE01}
Plenge, F., Rodin, P., Sch{\"o}ll, E. and Krischer, K. (2001).
  \enquote{{Breathing current domains in globally coupled electrochemical
  systems: A comparison with a semiconductor model},} \emph{Phys.~Rev.~E}
  \textbf{64}, p. 056229.

\bibitem[{Pyragas(1992)}]{PYR92}
Pyragas, K. (1992). \enquote{Continuous control of chaos by self-controlling
  feedback,} \emph{Phys. Lett.~A} \textbf{170}, p. 421.

\bibitem[{Residori(2005)}]{Residori2005201}
Residori, S. (2005). \enquote{{Patterns, fronts and structures in a
  Liquid-Crystal-Light-Valve with optical feedback},} \emph{Phys. Rep.}
  \textbf{416}, 5-6, p. 201–272.

\bibitem[{Sakurai \emph{et~al.}(2002)Sakurai, Mihaliuk, Chirila and
  Showalter}]{sakurai2002design}
Sakurai, T., Mihaliuk, E., Chirila, F. and Showalter, K. (2002).
  \enquote{Design and control of wave propagation patterns in excitable media,}
  \emph{Science} \textbf{296}, 5575, pp. 2009--2012.

\bibitem[{Schimansky-Geier \emph{et~al.}(2007)Schimansky-Geier, Fiedler, Kurths
  and Sch{\"o}ll}]{schimansky2007analysis}
Schimansky-Geier, Fiedler, B., Kurths, J. and Sch{\"o}ll, E. (eds.) (2007).
  \emph{Analysis and control of complex nonlinear processes in physics,
  chemistry and biology} (World Scientific, Singapore).

\bibitem[{Schimansky-Geier \emph{et~al.}(1983)Schimansky-Geier, Mikhailov and
  Ebeling}]{schimanskygeier1983effect}
Schimansky-Geier, L., Mikhailov, A.~S. and Ebeling, W. (1983). \enquote{Effect
  of fluctuation on plane front propagation in bistable nonequilibrium
  systems,} \emph{Ann. Phys. (Leipzig)} \textbf{495}, 4-5, pp. 277--286.

\bibitem[{Schlesner \emph{et~al.}(2008)Schlesner, Zykov, Brandtst{\"a}dter,
  Gerdes and Engel}]{schlesner2008efficient}
Schlesner, J., Zykov, V., Brandtst{\"a}dter, H., Gerdes, I. and Engel, H.
  (2008). \enquote{Efficient control of spiral wave location in an excitable
  medium with localized heterogeneities,} \emph{New J. Phys.} \textbf{10}, 1,
  p. 015003.

\bibitem[{Schlesner \emph{et~al.}(2006)Schlesner, Zykov, Engel and
  Sch{\"o}ll}]{schlesner2006stabilization}
Schlesner, J., Zykov, V., Engel, H. and Sch{\"o}ll, E. (2006).
  \enquote{Stabilization of unstable rigid rotation of spiral waves in
  excitable media,} \emph{Phys. Rev. E} \textbf{74}, 4, p. 046215.

\bibitem[{Schl{\"o}gl(1972)}]{Schlogl1972crm}
Schl{\"o}gl, F. (1972). \enquote{Chemical reaction models for non-equilibrium
  phase transitions,} \emph{Z. Phys. A} \textbf{253}, 2, pp. 147--161.

\bibitem[{Schl\"ogl \emph{et~al.}(1983)Schl\"ogl, Escher and
  Berry}]{Schloegl1983fluctuations}
Schl\"ogl, F., Escher, C. and Berry, R.~S. (1983). \enquote{Fluctuations in the
  interface between two phases,} \emph{Phys. Rev. A} \textbf{27}, pp.
  2698--2704.

\bibitem[{Schneider \emph{et~al.}(2009)Schneider, Sch{\"o}ll and
  Dahlem}]{SCH09c}
Schneider, F.~M., Sch{\"o}ll, E. and Dahlem, M.~A. (2009).
  \enquote{{Controlling the onset of traveling pulses in excitable media by
  nonlocal spatial coupling and time delayed feedback},} \emph{Chaos}
  \textbf{19}, p. 015110.

\bibitem[{Sch{\"o}ll(1986)}]{SCH86a}
Sch{\"o}ll, E. (1986). \enquote{Influence of boundaries on dissipative
  structures in the {S}chl{\"o}gl model,} \emph{Z.~Phys.~B} \textbf{62}, p.
  245.

\bibitem[{Sch{\"o}ll(1987)}]{SCH87}
Sch{\"o}ll, E. (1987). \emph{Nonequilibrium Phase Transitions in
  Semiconductors} (Springer, Berlin).

\bibitem[{Sch{\"o}ll(2001)}]{SCH01}
Sch{\"o}ll, E. (2001). \emph{{Nonlinear spatio-temporal dynamics and chaos in
  semiconductors}} (Cambridge University Press, Cambridge), {Nonlinear Science
  Series}, Vol. 10.

\bibitem[{Sch{\"o}ll(2010)}]{SCH09}
Sch{\"o}ll, E. (2010). \enquote{Pattern formation and time-delayed feedback
  control at the nano-scale,} in G.~Radons, B.~Rumpf and H.~G. Schuster (eds.),
  \emph{Nonlinear Dynamics of Nanosystems} (Wiley-VCH, Weinheim), ISBN
  978-3-527-40791-0, pp. 325--367.

\bibitem[{Sch{\"o}ll and Schuster(2008)}]{scholl2008handbook}
Sch{\"o}ll, E. and Schuster, H.~G. (eds.) (2008). \emph{Handbook of chaos
  control} (Wiley-VCH, Weinheim).

\bibitem[{Shima and Kuramoto(2004)}]{SHI04}
Shima, S.-I. and Kuramoto, Y. (2004). \enquote{Rotating spiral waves with
  phase-randomized core in nonlocally coupled oscillators,} \emph{Phys. Rev.~E}
  \textbf{69}, 3, p. 036213.

\bibitem[{Siebert \emph{et~al.}(2014)Siebert, Alonso, B{\"a}r and
  Sch{\"o}ll}]{SIE14}
Siebert, J., Alonso, S., B{\"a}r, M. and Sch{\"o}ll, E. (2014).
  \enquote{Control of reaction-diffusion patterns by asymmetric nonlocal
  coupling,} arXiv:1401.3111, \urlprefix\url{http://arxiv.org/abs/1401.3111}.

\bibitem[{Tr{\"o}ltzsch(2010)}]{troltzsch2010optimal}
Tr{\"o}ltzsch, F. (2010). \emph{Optimal control of partial differential
  equations: theory, methods, and applications}, Vol. 112 (American
  Mathematical Society, Providence).

\bibitem[{Vanag and Epstein(2008)}]{vanag2008design}
Vanag, V. and Epstein, I. (2008). \enquote{Design and control of patterns in
  reaction-diffusion systems,} \emph{Chaos} \textbf{18}, 2, pp. 026107--026107.

\bibitem[{Wolff(2002)}]{wolff2002lokale}
Wolff, J. (2002). \emph{Lokale Kontrolle der Musterbildung bei der CO-Oxidation
  auf einer Pt (110)-Oberfl{\"a}che}, Ph.D. thesis.

\bibitem[{Wolff \emph{et~al.}(2003{\natexlab{a}})Wolff, Papathanasiou,
  Rotermund, Ertl, Katsoulakis, Li and Kevrekidis}]{wolff2003wave}
Wolff, J., Papathanasiou, A., Rotermund, H., Ertl, G., Katsoulakis, M., Li, X.
  and Kevrekidis, I. (2003{\natexlab{a}}). \enquote{Wave initiation through
  spatiotemporally controllable perturbations,} \emph{Phys. Rev. Lett.}
  \textbf{90}, 14, p. 148301.

\bibitem[{Wolff \emph{et~al.}(2001)Wolff, Papathanasiou, Kevrekidis, Rotermund
  and Ertl}]{wolff2001spatiotemporal}
Wolff, J., Papathanasiou, A.~G., Kevrekidis, I.~G., Rotermund, H.~H. and Ertl,
  G. (2001). \enquote{Spatiotemporal addressing of surface activity,}
  \emph{Science} \textbf{294}, 5540, pp. 134--137.

\bibitem[{Wolff \emph{et~al.}(2003{\natexlab{b}})Wolff, Papathanasiou,
  Rotermund, Ertl, Li and Kevrekidis}]{wolff2003gentle}
Wolff, J., Papathanasiou, A.~G., Rotermund, H.~H., Ertl, G., Li, X. and
  Kevrekidis, I.~G. (2003{\natexlab{b}}). \enquote{Gentle dragging of reaction
  waves,} \emph{Phys. Rev. Lett.} \textbf{90}, 1, p. 018302.

\bibitem[{Zel'dovich and Frank-Kamenetskii(1938)}]{zeldovich1938theory}
Zel'dovich, Y.~B. and Frank-Kamenetskii, D.~A. (1938). \enquote{On the theory
  of uniform flame propagation,} \emph{Dokl. Akad. Nauk SSSR} \textbf{19}, pp.
  693--798.

\bibitem[{Zykov and Engel(2004)}]{zykov2004feedback}
Zykov, V. and Engel, H. (2004). \enquote{Feedback-mediated control of spiral
  waves,} \emph{Physica D} \textbf{199}, 1, pp. 243--263.

\bibitem[{Zykov \emph{et~al.}(2004)Zykov, Bordiougov, Brandtst\"adter, Gerdes
  and Engel}]{zykov2004global}
Zykov, V.~S., Bordiougov, G., Brandtst\"adter, H., Gerdes, I. and Engel, H.
  (2004). \enquote{Global control of spiral wave dynamics in an excitable
  domain of circular and elliptical shape,} \emph{Phys. Rev. Lett.}
  \textbf{92}, p. 018304.

\end{thebibliography}

\end{document}